\PassOptionsToPackage{hyphens}{url}
\PassOptionsToPackage{dvipsnames}{xcolor}
\documentclass[format=acmsmall, review=false]{acmart}
\usepackage{acm-ec-26-proc}
\usepackage{booktabs} 
\usepackage[ruled]{algorithm2e} 
\usepackage{natbib} 
\usepackage{enumitem}

\SetAlFnt{\small}
\SetAlCapFnt{\small}
\SetAlCapNameFnt{\small}
\SetAlCapHSkip{0pt}
\IncMargin{-\parindent}

\setcitestyle{authoryear}

\usepackage{array}
\usepackage{nicefrac}
\usepackage{graphicx}
\usepackage{tikz}
\usepackage{wrapfig}
\usetikzlibrary{decorations.pathreplacing}

\usepackage{xcolor}
\usepackage{pifont}
\usepackage{mdframed}
\usepackage{etoolbox}
\usepackage{colortbl}
\usepackage{thmtools}
\usepackage{thm-restate}
\usepackage[sort&compress,nameinlink]{cleveref}

\definecolor{darkblue}{rgb}{0.0, 0.0, 0.4}
\definecolor{winered}{rgb}{0.5,0.1,0.1}
\definecolor{darkgreen}{rgb}{0,0.35,0}
\definecolor{verylightgray}{gray}{0.9}
\definecolor{crimson}{RGB}{220, 20, 60}

\allowdisplaybreaks
\renewcommand\emph[1]{{\color{winered}\textit{#1}}}

\newcolumntype{Y}{>{\centering\arraybackslash}X}
\newcolumntype{Z}{>{\raggedleft\arraybackslash}X}
\newcolumntype{a}{>{\columncolor{black!05}}Y}

\newcommand*{\medcup}{\mathbin{\scalebox{1.5}{\ensuremath{\cup}}}}

\usepackage{comment}

\usepackage{mathtools}

\renewcommand*{\leq}{\leqslant}

\renewcommand*{\geq}{\geqslant}
\renewcommand{\epsilon}{\varepsilon}

\newcommand{\naturals}{{\mathbb{N}}}
\newcommand{\reals}{{\mathbb{R}}}

\newcommand{\calF}{{\mathcal{F}}}

\newcommand{\vclaim}{{\mathrm{claim}}}

\newcommand{\avgsat}{{\mathrm{avgsat}}}

\renewcommand{\part}{{\mathrm{part}}}

\newcommand{\lv}[1]{}
\newcommand{\sv}[1]{#1}

\Crefname{table}{Table}{Tables}
\Crefname{figure}{Figure}{Figures}
\Crefname{theorem}{Theorem}{Theorems}
\Crefname{definition}{Definition}{Definitions}
\Crefname{corollary}{Corollary}{Corollaries}
\Crefname{observation}{Observation}{Observations}
\Crefname{question}{Question}{Questions}
\Crefname{lemma}{Lemma}{Lemmas}
\Crefname{example}{Example}{Examples}
\Crefname{reduction}{Reduction}{Reductions}
\Crefname{construction}{Construction}{Constructions}
\Crefname{proposition}{Proposition}{Propositions}
\Crefname{lstlisting}{Listing}{Listings}
\Crefname{equation}{Relation}{Relations}
\Crefname{claim}{Claim}{Claims}

\theoremstyle{definition}
\newtheorem{definition}{Definition}
\newtheorem{example}{Example}

\theoremstyle{plain}
\newtheorem{theorem}{Theorem}

\newtheorem{claim}[theorem]{Claim}
\newtheorem{proposition}[theorem]{Proposition}

\ccsdesc[500]{Theory of computation~Algorithmic game theory and mechanism design}
\ccsdesc[500]{Applied computing~Voting / election technologies}

\keywords{Multiwinner Elections,
Positive and Negative Votes,
Proportionality}

\copyrightyear{2026}
\acmYear{2026}
\setcopyright{cc}
\setcctype{by}
\acmConference[EC '26]{The 27th ACM Conference on Economics and Computation}{July 06--10}{Rome, Italy}
\acmBooktitle{The 27th ACM Conference on Economics and Computation (EC '26), July 06--10, 2026, Rome, Italy}
\acmDOI{10.1145/3821539.3827792}
\acmISBN{979-8-4007-2813-6/2026/07}

\title[Proportional Committee Elections with Positive and Negative Votes]{Proportional Committee Elections\\ with Positive and Negative Votes}

\author{Sonja Kraiczy}
\affiliation{
  \institution{University of Oxford}
  \city{Oxford}
  \country{United Kingdom}
}

\author{Georgios Papasotiropoulos}
\affiliation{
  \institution{University of Warsaw}
  \city{Warsaw}
  \country{Poland}
}

\author{Grzegorz Pierczyński}
\affiliation{
  \institution{University of Warsaw}
  \city{Warsaw}
  \country{Poland}
}

\author{Piotr Skowron}
\affiliation{
  \institution{University of Warsaw}
  \city{Warsaw}
  \country{Poland}
}

\begin{abstract}
In the classic committee election setting each voter approves a subset of candidates and the goal is to select $k$ winners based on these preferences. A central focus of recent research in the area has been to achieve proportional representation. In this work, we explore notions of proportionality in a more expressive setting that allows voters to vote against candidates---a common feature on online polling platforms and beyond. We propose two conceptually distinct interpretations of negative votes, resulting in different perspectives of proportionality. In the first, preventing the election of disapproved candidates is as important to voters as electing approved ones. In the second, approvals and disapprovals are treated separately, with each receiving its own fairness guarantees. For each approach, we introduce suitable axioms capturing proportionality and examine their satisfiability by appropriate variants of Phragmén's rule, Proportional Approval Voting rule (PAV), and the Method of Equal Shares (MES).
\end{abstract}

\begin{document}

\maketitle

\section{Introduction}

Consider the problem of selecting up to $k$ winners among a set of $m$ candidates based on voters' preferences. Voters can \emph{approve} or \emph{disapprove} a candidate, or otherwise \emph{abstain}. The objective is to select a set of winning candidates in a \emph{proportional} manner, which roughly means that \textit{each group of voters with sufficiently similar preferences should have an influence on the outcome in proportion to its size}. 
In essence, proportionality is a notion of fairness that captures the idea that groups of like-minded voters should obtain representation that reflects their relative size.
In particular, a group of voters with similar preferences that constitutes an $\alpha$-fraction of the electorate deserves an $\alpha$-fraction share of the committee.

In recent years, the proportional selection problem for approval votes has become one of the most active research directions and core challenges in the field of computational social choice~\citep{lac-sko:multiwinner-book,FSST-trends,pet-sko:laminar}.
In the classic approval voting framework, the design of allowed votes lacks the nuance to capture whether not approving a candidate corresponds to indifference or disapproval towards that candidate.
This limitation can be addressed by allowing voters to explicitly disapprove of candidates.
Incorporating negative votes has been identified as an important open problem in the field \cite{brill2023algorithms,lac-sko:multiwinner-book,rey2023computational}.
Indeed, it introduces significant complexity, the primary conceptual challenge being to capture proportionality in the presence of negative votes. To make this explicit, we consider the following instance.

\begin{example}
\label{ex:intro_example}
Consider a committee election with positive and negative votes where there are $27$ candidates, divided into three groups, namely $C_1$, $C_2$, and $C_3,$
and $100$ voters, divided into two disjoint groups, namely $V_1$ and $V_2$. The preferences of the voters over the candidates' sets
appear below and say for simplicity that we break ties lexicographically. 
\smallskip

%
{\centering    \begin{tabular}{>{\bfseries}c c c c}
\toprule
& $\boldsymbol{C_1}$\textbf{ (10 candidates)} & $\boldsymbol{C_2}$\textbf{ (7 candidates)} & $\boldsymbol{C_3}$\textbf{ (10 candidates)} \\
\midrule
$\boldsymbol{V_1}$\textbf{ (60 voters)} & neutral & approve ({\color{darkgreen}\ding{51}}) & approve ({\color{darkgreen}\ding{51}}) \\
$\boldsymbol{V_2}$\textbf{ (40 voters)} & approve ({\color{darkgreen}\ding{51}}) & disapprove ({\color{winered}\ding{55}}) & neutral \\
\bottomrule
\end{tabular}
\par}

\smallskip
Say that the size of the committee to be elected is $k=10$.
If we ignore the negative votes and apply the na\"ive utilitarian rule that simply selects the $k$ candidates with the highest approval counts, we would end up choosing candidates exclusively from $C_2 \cup C_3$. 
This would be unfair, as it would effectively disregard the preferences of the voters in $V_2$ who do not approve any elected candidate. 
In particular, under the specified tie-breaking order, every candidate from $C_2$ will be elected, so, not only would the voters in $V_2$ fail to elect any of their preferred candidates, but also the selected outcome includes all the candidates they oppose. 
Evidently, while all candidates in $C_2\cup C_3$ are approved by the voters in $V_1$, selecting candidates from $C_2$ causes dissatisfaction among the voters in $V_2$ and, hence, candidates from $C_3$ appear to be a better fit than those from $C_2$. 
A greedy rule that selects candidates in order of net approval (i.e., the number of approvals minus the number of disapprovals) would elect $C_3$. 
This solution though again leaves $40\%$ of the population without representation.\looseness-1

How many candidates from $C_3$ should then be elected? This depends on how we interpret voters' disapproval statements. One could argue that electing candidates from $C_3$ instead of $C_2$ should not affect the entitlement of group $V_2$ to $4$ candidates from $C_1$. On the other hand, one might say that $V_2$ gains satisfaction from blocking the election of candidates they dislike and should ``pay'' for this influence. Under this view, a proportional outcome might include fewer than $4$ candidates from $C_1$. These two intuitions reflect the two distinct models of proportionality we propose and examine.
\hfill $\lrcorner$
\end{example}

\subsection{Negative Votes in the (Real) World of Collective Decision-Making}
\label{sec:applications}
There is significant evidence that negative voting influences voting behavior and is often desired in practice. 
A study on the 2020 U.S. Presidential election found that approximately a third of American voters cast their votes \textit{more ``against'' a candidate than ``for'' one}~\citep{gar-fer:negative-voting-us-campaign}.
Another example which highlights the importance of incorporating negative feedback comes from \emph{participatory budgeting (PB)}~\cite{rey2023computational}, a form of citizen participation employed in cities where citizens directly decide how a portion of a public budget is spent. While some cities, like Madrid~\cite{wikipblink}, allow voters to express opposition to projects, most PB elections do not.
This has led to issues, as seen, for instance, when residents of a large Warsaw housing estate protested against the construction of a playground selected through the PB process~\citep{pb-protests}. Since the formal model of PB generalises committee elections, our work is a fundamental step towards incorporating negative voting in the PB model while maintaining guarantees on proportional representation. Indeed, some of our proposed rules can already be applied in PB. 

\emph{Public engagement platforms} (e.g., \href{https://pol.is/}{\texttt{pol.is}}) allow citizens to safely exchange ideas with others, authorities, experts, and journalists. Participants can provide open-ended feedback and agree or disagree with statements, highlighting both common ground and differences while helping shape the political or news agenda. For instance, such systems have been used to explore issues like universal basic income, taxation, biodiversity policies in New Zealand~\citep{hive}, and multiple economic, technological, and social regulatory policies in Taiwan~\citep{taiwan1,taiwan2}, demonstrating how positive and negative votes can support large-scale participatory deliberation.

Further examples where negative voting has been used in practice include \emph{Decentralized Autonomous Organizations} (e.g., \href{https://projectcatalyst.io}{\texttt{projectcatalyst.io}}
), where it is employed to decide which submitted proposals receive funding.
Moreover, in constitutional AI \citep{bai2022constitutional}, a \emph{constitution}---a slate of ethical principles---is used to align LLMs with human values.
For this purpose, Anthropic and the \href{https://www.cip.org}{Collective Intelligence Project} recently used a civic participation approach to draft a constitution based on the preferences of approximately 1000 Americans. The process involved participants voting on ethical principles by approving, disapproving or abstaining, which were later aggregated to form the
final constitution \citep{anthropic2024}. 

Lastly, many are familiar with negative voting from social media platforms such as Reddit, or Stack Exchange, where users employ upvotes and downvotes to express approval or disapproval and capture public sentiment, shaping both content visibility and collective opinion.

While positive and negative votes have a wide range of applications, the methods used to aggregate voters' preferences and reach a collective decision remain rather limited in real-life cases. In particular, in many scenarios the greedy rule based on the net approval scores of candidates is employed. We refer back to \Cref{ex:intro_example} for an illustration of why this approach is not satisfactory from a fairness perspective.
More precisely, on Pol.is the net agreement score is used to identify consensus statements and to rank proposals or ideas. Net approval scores are also used to rank answers, comments and projects on Stack Exchange, Reddit and Project Catalyst, respectively. 
Finally, several small-scale PB tools have experimented with the option of allowing for negative votes. For example, the only edition of participatory budgeting in Poland (in Sopot) that allowed for negative votes also selected projects based on their net approval scores.

To sum up, negative votes are not only a natural format for expressing voters' preferences, but are also already used in practice. Yet, they are typically combined with simple greedy aggregation methods, providing a clear call to action for the development of voting rules that offer proportional representation guarantees.

\subsection{Contribution of the Paper}

Our main conceptual contribution consists in proposing two (incomparable) approaches to defining proportionality in the presence of negative votes. The two approaches differ in how (and whether) the utilities from electing preferred candidates and from blocking the undesirable ones compare. We introduce suitable proportionality axioms for each, and we study voting rules that generalize classic ones, from the approval setting, specifically \emph{Phragm\'en's rule}, \emph{Proportional Approval Voting rule (PAV)}, and the \emph{Method of Equal Shares (MES)}. Finally, we provide positive and negative results on these rules concerning the formulated axioms. 
Our analysis employs several new techniques that are potentially applicable in broader contexts, e.g., in the presence of diversity constraints.  

In our first interpretation of negative votes, 
not selecting a disapproved candidate is considered just as important to a voter as selecting an approved one.
We call this the \emph{symmetric utility model}. 
Here, for each candidate $c$ we introduce its virtual negative counterpart $\neg c$, and impose constraints to ensure that $c$ and $\neg c$ are never selected together. Each voter's utility is additive, derived either from including $c$ in the winning committee if approved, or $\neg c$ if the voter disapproves of $c$. 
In this context, we introduce suitable proportionality axioms, generalizing established notions.
Specifically, we exploit the notion of \emph{$\ell$-cohesiveness}, which captures when a group $S$ deserves $\ell$.
This is the core idea of proportionality axioms such as \emph{Proportional Justified Representation} (PJR) \cite{sanchez2017proportional}, \emph{Extended Justified Representation} (EJR) \cite{aziz2017justified} and \emph{proportionality degree} \cite{skowron:prop-degree}.
We study a natural variant of Phragm\'en’s rule that applies in the examined setting, we show that it satisfies PJR and we provide an almost-tight guarantee for the proportionality degree of the rule. 
We then move to the PAV rule, and, as our main technical contribution we show that this rule gives near-optimal proportionality guarantees for $\ell$-cohesive groups of voters.
After examining the two rules and their proportionality guarantees, we identify a phenomenon that arises in them due specifically to the presence of negative votes. In particular, under both rules, voters may expend their influence trying to block candidates that have no chance of being elected, which can inadvertently reduce their effective impact on the outcome for candidates they support. We formalize this issue by introducing notions capturing outcome-relevant negative votes, and we take certain steps towards mitigating the observed problem, while preserving the proportional representation guarantees.
We finally position the examined symmetric utility setting as a special case of elections under general feasibility constraints \cite{masavrik2024generalised}, connecting our proportionality axioms to this broad recently introduced framework. In doing so, we highlight a certain limitation of the existing notions and offer a reformulation of them to address it; due to space constraints we defer these results to the full version of our work \cite{kraiczy2025proportionality}.

Our second interpretation of negative votes gives rise to a fundamentally different approach, called the \emph{asymmetric utility model}. The distinction stems from the view that voters deserve to be represented by members of the elected committee, regardless of whether they have already influenced the outcome by blocking the selection of certain disapproved candidates. Thus, we treat the right of a voter to be represented and their right to block candidates they oppose asymmetrically. Our axioms for this model are based on an adaptation of the idea of \emph{$\ell$-cohesiveness} in which a group's voting power is determined not merely by its size but by its \emph{effective size}---a measure that accounts for the number of opponents of the candidates the group supports.
Our positive results build on priceable voting rules, where the election can be viewed as a process where voters ``purchase'' candidates~\citep{pet-sko:laminar}.
It turns out that MES and Phragm\'en's rule can be adapted to satisfy our fairness objectives, by introducing the concept of a ``\emph{virtual tax}'' imposed by opponents of the candidates, which effectively increases the cost of disapproved candidates. 
In contrast, we show that there is no natural adaptation of PAV to this setting. 

\subsection{Related Work}

Proportionality in committee elections is a central topic in computational social choice \cite[Chapter 4]{lac-sko:multiwinner-book}. The election framework where voters can express both positive and negative votes towards candidates or abstain (often referred to as \emph{ternary} or \emph{trichotomous voting}) has been studied already \citep{yilmaz1999can,brams1983approval,alcantud2014dis,gonzalez2019dilemma,zhou2019parameterized,baumeister2015winner,brams1977advantageous,el2022comparing,laruelle2021not}. However, to our knowledge, the only existing work that merges both topics is by \citet{talmon2021proportionality}. There, the authors introduce seven proportionality axioms, which differ significantly from the ones proposed in our work. They further examine variants of the Monroe, Chamberlin-Courant, STV, and PAV voting rules, all adapted to the examined setting. None of these rules satisfy any of the considered axioms; instead the authors rely on simulations to examine how often the rules produce outcomes that align with their proportionality criteria. Therefore, our work is the first to provide theoretical guarantees for the examined problem.

Our problem can be viewed as a special case of \emph{elections with general feasibility constraints}, a setting in which the concept of Base proportionality has been recently explored. \citet{masavrik2024generalised} adapt PAV and Phragm\'en to that framework and show these rules satisfy desirable proportionality axioms provided the constraints form \emph{matroids}.
These results do not apply to our setting, as our feasibility constraints do not have a matroid structure. 
The setting we examine can also be viewed as a special case of voting under \emph{weak preference orders} or \emph{cardinal preferences} (cf. \citep{aziz2020expanding,pet-pie-sko:c:participatory-budgeting-cardinal}). 
A related line of research is that of \emph{public decisions}, where voters submit votes as in trichotomous voting and a decision must be made for each of a number of issues. Proportionality has also been studied in this framework \cite{fre-kah-pen:variable_committee_elections,sko-gor:proportional_public_decisions,bri-mar-pap-pet:proporitonlaity-on-independent-issues}.
A key distinction is that in the model of public decisions there is no upper bound on the number of selected candidates; this is enough to make the two frameworks critically different, both conceptually and technically.

\section{Preliminaries}
\label{sec:prelims}

An \emph{election} $E$ is a tuple $(C, V, k, B)$, where $C := \{c_1, c_2, \ldots, c_m\}$ is the set of candidates, $V := \{1, 2, \ldots, n\}$ is the set of voters, $k \leq m$ is an integer corresponding to an upper bound on the number of candidates to be elected, and $B:=(B_i)_{i \in V},$ with $B_i$ being the ballot submitted by the $i$-th voter. Since voters are allowed to express both positive and negative preferences for candidates, each ballot $B_i$ is represented as a pair $B_i = (A_i, D_i)$, with $A_i, D_i \subseteq C$ and $A_i \cap D_i = \emptyset$. $A_i$ denotes the set of candidates \emph{approved by} voter $i$, and $D_i$ is the set of candidates \emph{disapproved/vetoed by} this voter. If a candidate $c \notin A_i \cup D_i$, we say that voter $i$ is \emph{indifferent} or \emph{neutral} toward $c$.

For each candidate $c\in C$, let $A_c$ and $D_c$ denote the sets of voters who approve and disapprove $c$, respectively. Formally, $A_c := \{i \in V: c \in  A_i\}$ and $D_c := \{i \in V: c \in  D_i\}$.
For each group of voters $S\subseteq V,$ let 
$A_S$ and $D_S$
denote the sets of candidates commonly approved and disapproved by $S,$ respectively. Formally, $A_S := \bigcap_{i\in S} A_i$ and $D_S := \bigcap_{i\in S} D_i$.
An election rule is a function $\mathcal{R}$ which for each election $E$ returns a set of at most $k$ candidates: a \emph{winning outcome/committee}. 
Clearly if $D_i=\emptyset$ for every voter $i\in V,$ the considered election is a classic \emph{approval-based committee election}. 

Below we describe election rules that are known to perform particularly well in terms of proportionality in the approval setting~\citep{lac-sko:multiwinner-book}. In the subsequent sections, we focus on designing generalizations of these rules that also perform well in the presence of negative votes.

\begin{description}
[leftmargin=2pt,topsep=4pt,itemsep=5pt]
\item[Phragm\'en's rule \textnormal{\citep{Phra94a,Janson16arxiv}}.] Each candidate is assumed to have a unit cost. Each voter has a virtual budget starting at 0 and increasing continuously over time. At time $t$, each voter has been allocated a total budget of $t$. When a group of voters approving a common candidate accumulates enough budget to cover the candidate's cost, the candidate is added to the winning committee and the budgets of the voters involved in the purchase are reset to 0. This process continues until $k$ candidates are selected. 
        \item[Thiele rules 
    \textnormal{\citep{Thie95a,Janson16arxiv}}.]
    Each rule in the class of Thiele methods is parameterized by a nondecreasing function $f:\{0,1,\dots,k\}\to \mathbb{R_{\geq \text{0}}}$ with $f(0)=0$. A voter $i$ assigns a score of
    $f(|W\cap A_i|)$ to a committee $W$. The winning outcome is the committee of size $k$ that maximizes $\sum_{i\in V}f(|W\cap A_i|).$ 
    \textbf{Proportional Approval Voting (PAV) rule} \citep{aziz2017justified, AEHLSS18, pet-sko:laminar} is the rule in the class of Thiele methods for which $f(x)=\sum_{j=1}^{x}\nicefrac{1}{j}.$ 
	\item[Method of Equal Shares (MES) \textnormal{\citep{pet-sko:laminar}}. ] The rule works in rounds.
	Let $b_i$ be the virtual budget of voter $i$, initially set to $\nicefrac{k}{n}$. Each candidate is assumed to have a unit cost.
	In each round, we consider every not yet elected candidate $c$ whose supporters have at least a total budget that suffices to buy $c$.
We say that such a candidate $c$ is \textit{$\rho$-affordable} for $\rho \in \mathbb{R}_{+}$ if
		$\sum_{i \in A_c} \min\left(b_i,\rho\right) =1.$
	The candidate that is $\rho$-affordable for the smallest value of $\rho$ is selected and added to the committee. The budgets of her supporters are then updated accordingly, i.e., $b_i := b_i - \min\left(b_i, \rho\right)$.
	The rule stops if there is no $\rho$-affordable candidate for any value of $\rho$. 
\end{description}

\section{The Symmetric Utility Model}
\label{sec:uniform}
In the \textit{symmetric utility model}, a voter derives one unit of utility both from including an approved candidate in the winning committee and from excluding a disapproved candidate from it. To capture this interpretation, we introduce \emph{virtual negative candidates}: for every \(c\in C\), we add a counterpart \(\neg c\). We reserve the term \emph{committee} for the set \(W\subseteq C\) of elected positive candidates, with \(|W|\leq k\). An \emph{outcome}, denoted by \(P\), may additionally contain virtual negative candidates, where selecting \(\neg c\) represents rejecting candidate \(c\). Feasibility requires that \(P\) contains at most one of \(c\) and \(\neg c\) for every \(c\in C\), and that \(P\cap C\) is a committee.

Note that if neither \(c\) nor \(\neg c\) is selected in an outcome, then adding \(\neg c\) preserves feasibility and weakly increases every voter's utility. Hence, without loss of generality, we may restrict attention to complete outcomes, in which exactly one of \(c\) and \(\neg c\) is selected for every candidate \(c\). Consequently, every committee \(W\subseteq C\) naturally induces the complete outcome
$P(W)=W\cup \{\neg c\mid c\notin W\}.$
The term \emph{symmetric utility model} refers to the symmetry in the voters' utilities: the utility derived from an approved candidate being included in \(W\) is identical to the utility derived from a disapproved candidate being excluded from \(W\). 
For each set $X \subseteq C$ we define {$\neg X := \{\neg c\mid c\in X\}$.

In this section, we show that straightforward extensions of well-established approval-based rules---namely, Phragm\'en's rule and PAV---to the augmented candidate set $C\cup \neg C$, subject to the feasibility constraints described above, continue to provide strong proportionality guarantees in the symmetric utility model.

At first sight, one might hope to obtain such guarantees directly from existing results. Indeed, natural extensions of Phragm\'en's rule and PAV are known to provide good guarantees in the simpler setting without an upper bound on the number of positive candidates to be selected; see, e.g., \cite{fre-kah-pen:variable_committee_elections,sko-gor:proportional_public_decisions}. Moreover, later work by \citet{masavrik2024generalised} shows that these extensions remain well behaved when the feasibility constraints form a matroid, as is the case, for instance, in standard committee elections and above mentioned variable-winner setting.
However, the feasibility constraints induced by the symmetric utility model are not matroid constraints. This distinction is important, since \citet{masavrik2024generalised} demonstrate limitations of such rules beyond matroid constraints, showing, for example, that PAV may fail proportionality in general non-matroid settings. Our contribution should not be seen as the definition of new voting rules but in developing a new technique that establishes positive proportionality guarantees for natural and well-established rules in a setting where existing matroid-based results do not apply.

We begin by introducing the axiomatic notion used to state our proportionality guarantees. Consider a group \(S\subseteq V\), which forms a \(\nicefrac{|S|}{n}\)-fraction of the electorate. The proportional claim of \(S\) is based on the candidates on which all voters in \(S\) agree. In the symmetric utility model, such common agreement can contribute to the group's satisfaction in two ways: by electing candidates commonly approved by \(S\), and by rejecting candidates commonly disapproved by \(S\). Since at most \(k\) candidates can be elected, the contribution from commonly approved candidates is capped at \(\min(k,|A_S|)\). Hence, the total common satisfaction on which the claim of $S$ can be based is
\(
    |D_S|+\min(k,|A_S|).
\)
We say that \(S\) is cohesive if it can claim a proportional share of this quantity, according to the following definition.
\begin{definition}
    \label{def:axiom}
Consider an election $E$ and let $S$ be a group of voters.
We say that $S$ is $\ell$-cohesive if $\ell\leq \nicefrac{|S|}{n}(|D_S|+\min(k,|A_S|))$, and that it is strongly $\ell$-cohesive if $\ell\leq \nicefrac{|S|}{n} \cdot \min(k,|A_S| + |D_S|)$.
\hfill $\lrcorner$\end{definition}

This notion is deliberately more conservative than the standard notion of
\(\ell\)-cohesiveness from approval-based committee elections. The reason is
that, in the examined model, proportionality claims must remain jointly
satisfiable even when candidates approved by one group are disapproved by
another. To see this, suppose that a group \(S\) approves a set \(A_S\) of
candidates, while the remaining voters disapprove exactly these candidates and
have no other approvals or disapprovals. Then a proportional outcome should
select roughly a \(\nicefrac{|S|}{n}\)-fraction of the candidates in \(A_S\)
and reject the remaining fraction, so that both groups receive their
proportional shares. Hence, even when \(D_S=\emptyset\), the claim of \(S\)
cannot in general coincide with the standard approval-based claim: increasing
it substantially would make the claims of the two groups impossible to satisfy
simultaneously. 
In other words, the condition is weaker compared to the approval setting, because it accounts for possible disapprovals that might appear among other voters.

The strong variant is more restrictive. It treats the commonly approved and
commonly disapproved candidates of \(S\) as a single pool of common claims and
caps this pool by \(k\). This cap should not be interpreted as an upper bound
on the utility a voter can obtain from rejected candidates, but rather as a simple strengthening allowing to treat positive and negative candidates symmetrically.

Given a group $S$ and a committee $W$, we define the utility of a voter $i \in S$ from $W$ as
$u_i(W) := |A_i \cap W| + |\neg D_i \cap W|$, and the average satisfaction of voters from $S$ as $\avgsat_S(W) := \nicefrac{1}{|S|} \cdot \textstyle\sum_{i \in S}u_i(W)$.

\subsection{Phragm\'en's Rule}
\label{sec:phragmen}
Recall the definition of Phragm\'en's rule for approval-based committee elections from \Cref{sec:prelims}. 
A natural first approach to extend it to the setting of negative votes could be as follows. Run two parallel elections: one (positive) election for positive candidates and one (negative) election for negative ones, with money accumulating at the same rate in both. When a candidate is elected in the positive election, their corresponding negative candidate is removed from the negative election and vice versa. This approach can fail to recognize synergies between groups that derive satisfaction differently,
leading to suboptimal outcomes. 
Indicatively, consider an election with $n=k=2\ell$ and voters being split into two equal groups, $S_1$ and $S_2$. There are four disjoint candidate sets $C_1, C_2, C_3, C_4$ with $|C_1| = |C_3| = 2\ell$ and $|C_2| = |C_4| = \ell$. Voters in $S_1$ collectively approve $C_1$, and each voter in $S_1$ votes negatively for a unique candidate from $C_4$. $S_2$ collectively disapproves $C_3$, and each voter in $S_2$ approves a unique candidate from $C_2$. 
According to the discussed variant we only select $\ell$ candidates from $C_1$ and $C_2$ and the negative counterparts of $\ell$ candidates from $C_3$ and all from $C_4$.
Any candidates not explicitly mentioned to have been selected remain unelected.
A solution that would better satisfy all voters is to select all candidates from $C_1$ along with the negative counterparts from $C_3$. 
In response, we adapt the 
rule as follows:

\medskip

\begin{mdframed}[linecolor=black, linewidth=0.3mm, innerleftmargin=10pt, leftmargin=10pt, rightmargin=10pt
]
\textbf{Phragm\'en's rule for positive and negative votes.} All voters continuously earn money at the same rate. They can spend it not only to elect candidates but also to block those they oppose. If a group of voters collectively accumulates enough funds to veto a (yet unelected) disapproved candidate, they do so by purchasing the corresponding negative candidate and their budget is reset. Electing a candidate $c$ costs one unit and removes $\neg c$ from consideration. Similarly, purchasing $\neg c$ for one unit removes $c$. This process continues until, for every original candidate \(c\), the outcome contains either \(c\) or its negative counterpart \(\neg c\), or until selecting any further positive candidate would violate the bound of \(k\) positive candidates.
\end{mdframed}
\medskip
Phragm\'en's rule does not satisfy EJR for the committee setting, hence our extension is not intended to satisfy its extension in the setting with positive and negative votes. Therefore, we focus on two other (weaker) properties concerning $\ell$-cohesive groups. The first of them is the \emph{Proportional Justified Representation} (PJR) axiom \cite{sanchez2017proportional,masavrik2024generalised}. In contrast to EJR, it only ensures that the group as a whole is adequately represented, without guaranteeing that any individual benefits sufficiently. 
The second one is the high \emph{proportionality degree}, guaranteeing that each $\ell$-cohesive group of voters will have high representation on average, again, relative to what they deserve \citep{sanchez2017proportional}.

\begin{restatable}{theorem}{thmPJR}\label{bpjrphragmen}
Let $W$ be an outcome returned by Phragm\'en's rule. For each strongly $\ell$-cohesive set of voters $S$ the following conditions are satisfied:
    \begin{align*}
    (1) \qquad &\left|\big(\medcup_{i \in S} A_i  \cup \medcup_{i \in S} \neg D_i\big) \cap W\right| \geq \ell, \qquad &&\text{(PJR)}\\
    (2) \qquad &\avgsat_S(W) \geq \frac{\ell-1}{2}. &&\text{(proportionality degree)}
    \end{align*}

\end{restatable}

Combining the guarantees from \Cref{bpjrphragmen} with the rule’s polynomial runtime makes it especially appealing when negative votes are present and the goal is to provide proportional representation to voters’ groups under the symmetric utility model.

\subsection{Proportional Approval Voting Rule}
\label{sec:pav}
A prominent rule for approval-based committee elections is Proportional Approval Voting (PAV). In contrast to Phragm\'en's rule, PAV extends straightforwardly to the model with negative votes. 
\medskip
\begin{mdframed}[linecolor=black, linewidth=0.3mm, innerleftmargin=10pt, leftmargin=10pt, rightmargin=10pt
]
\textbf{PAV rule for positive and negative votes.}
It selects a feasible outcome $W$ that maximizes:
\begin{align*}
\sum_{i \in V} f(|W \cap (A_i \cup \neg D_i)|), \qquad \text{where} \quad f(x) = \sum_{j = 1}^x\nicefrac{1}{j}.
\end{align*}
\end{mdframed}
\medskip
We prove that PAV provides a guarantee closely approximating $\ell$ on average for voters of every $\ell$-cohesive set $S$ (the smaller the size of $S$ is, the closer is the guarantee). This means that, even in the case where there is no single voter from $S$ having a utility of at least $\ell$ (as required by EJR), on average, the voters from $S$ have a utility close to $\ell$. Hence we once again establish a strong proportionality guarantee.

\begin{restatable}{theorem}{maintech}\label{thm:pav-negative-votes}
\label{thm:pavpositive}
    Consider an election $E$ and let $W$ be an outcome of PAV. For each $\ell$-cohesive $S\subseteq V$:
\begin{align*}
\avgsat_S(W) \geq (1 - \epsilon)\ell - \nicefrac{3}{2}, \qquad \text{where  } \,\,\epsilon := \frac{2}{k + 4} \text{.}
\end{align*}
\end{restatable}
\begin{proof}
    We begin by outlining the strategy of the proof. Assume, for the sake of contradiction, that there exists an outcome $W$ selected by PAV where the group of voters $S$ achieves a lower satisfaction than stated in the theorem. To address this, we consider a series of possible swaps that replace certain candidates in the committee $W$ with candidates outside $W$. Instead of analyzing each individual swap, we aggregate them to compute the overall change in the PAV-score of $W$ resulting from these swaps. It is crucial that these swaps are applied independently to the original committee $W$, rather than sequentially. It suffices to demonstrate that the aggregated change in the PAV-score is positive. Then, by the pigeonhole principle, this implies that there is at least one swap that can improve the PAV-score of the outcome $W$, leading to a contradiction. This line of reasoning was employed in the original proof that PAV satisfies EJR without negative votes. However, in the original setting, swaps simply involve replacing candidates one for one; an approach that fails in our setting. This is because adding a candidate $c$ to the committee may require not just removing an arbitrary candidate from $W$, but also addressing the presence of a ``virtual candidate'' $\neg c$, which represents the exclusion of $c$ from $W$. 
Hence, our analysis has to be much more nuanced, involving the consideration of swaps between groups of candidates rather than individual candidates, which makes it more technically demanding.

Let us now start by introducing some additional notation. 
Let $A^+$ be an arbitrary subset of $A_S$ of size $d^+ = \min(k,|A_S|)$. Let $A= \neg D_S \cup A^+$.
These are the candidates we will be trying to add to the committee $W$, while removing others to maintain feasibility. We denote the size of $A$ as $d$, thus:
\begin{align*}
d = |D_S| + \min(|A_S|, k) \text{.}
\end{align*}

For each voter $i$, we define 
$U_i := A_i \cup \neg D_i$.
Thus, $U_i$ represents the set of both positive and negative candidates that can contribute to the voter's satisfaction.
If $W < k$ then in order to add any candidate from $U_i \setminus W$ to $W$ we need to remove at most one candidate from $W$. Then the proof simplifies, and the reasoning from the proof of Theorem~6 by \citet{masavrik2024generalised} can be applied.

We split the committee $W$ into positive candidates $W^+$ and negative candidates $W^-$. We assume in the following that $|W^+|=k$; if not, the outcome can simply be extended by adding dummy candidates that are neither approved nor disapproved by any voter. Let $a=|A^+\cap W^+|$. We assume that $a<k$ as otherwise $W^+\subseteq A_S$ and the committee $W\cup \{\neg c\mid c\in D_S\}$ is feasible. This committee would already imply the desired result. Further, we assume without loss of generality that for every $c\in C$, either $c\in W^+$ or $\neg c \in W^-$. Indeed, if this is not the case, then $\neg c$ can be always added to the outcome, and the PAV score will not decrease after such a change. Finally, we set:
$k':=k-a=|W^+\setminus A^+| \text{.}
$

We will now define $k'$ injections from $A^+\setminus W^+$ to $W^+\setminus A^+$. Note that 
$|A^+\setminus W^+| = d^+ - a \leq k - a = |W^+\setminus A^+| \text{.}
$
For convenience we label the candidates in $A^+\setminus W^+$ as $a_0,\ldots, a_{d^+-a-1}$, and the candidates in $W^+\setminus A^+$ as $b_0,\ldots, b_{k'-1}$. We define $\phi_i$ for $i=0,\ldots,k'-1$ as $\phi_i(a_j):=b_{j+i \text{ mod }k'},\text{for } j=0,\ldots, d^+-a-1.
$

We begin by considering a concrete example of how to construct the exchanges from a single fixed injection $\phi$ before exhibiting our general construction. Let $k=3$ be the committee size, and let $W=\{c_1, c_2, c_3, \neg c_4,\neg c_5 ,\neg c_6\}$, i.e., $W^+=\{c_1, c_2, c_3\}$ and $W^-=\{\neg c_4, \neg c_5, \neg c_6\}$ Suppose that group $S$ jointly approves $A_S=\{c_4, c_5, c_6\}$ and $D_S = \{\neg c_1,\neg c_2\}$.
Then, we want to add elements from $A_S \cup D_S$ and remove elements from $W$. For feasibility, adding a negative candidate (i.e., one from $D_S$) only requires removing the corresponding positive. However, adding a positive candidate (i.e., one from $A_S$) requires removing both the corresponding negative and an additional positive from $W$, as determined by the injection.
Note that $W^+\cap A^+=\emptyset$ and consider a sample injection $\phi$ from $A^+$ to $W^+$: $\phi(c_4)=c_1$, $\phi(c_5)= c_2$ and $\phi(c_6)=c_3$. The primary swaps associated with $\phi$ are depicted in \Cref{fig:pavtables} (Left); to add $x$ we remove $\neg x$ and $\phi(x)$.
The set of swaps associated with the injection has the following property: if we restrict our attention to rows where positive candidates are being added, any positive candidate is removed in at most one such swap. Consequently, each positive candidate is removed in at most two swaps across the entire table.
However, for our proof we will need that each candidate (positive or negative) is either removed or added in at most one swap; and, clearly, any candidate might be considered for addition or removal but not both.
To achieve this, we iteratively identify pairs of swaps that involve the removal of the same positive candidate $x$. These pairs have the following structure: in one swap, $x$ is removed to make room for $\neg x$; in the other, $x$ and some $\neg y$ are removed to accommodate $y$. We merge these into a single exchange where $\neg x$ and $y$ are added, and $x$ and $\neg y$ are removed. The merged exchanges are illustrated in \Cref{fig:pavtables} (Right). It is evident that in the merged table every candidate is removed or added at most once, our desired property.
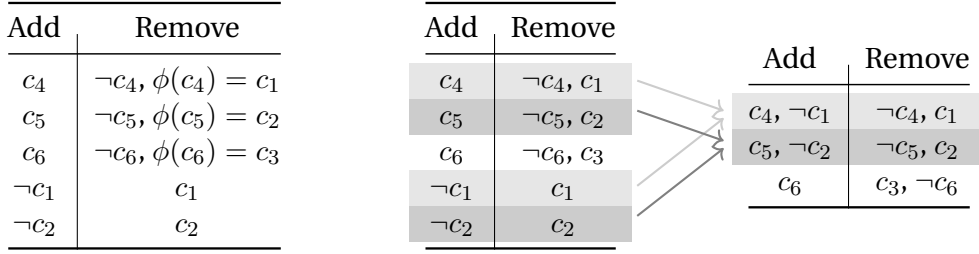
\begin{figure}
\begin{center}
    \begin{tikzpicture}
        \node (table1) at (-5,0) {
            \begin{tabular}{@{} c|c @{}}
                \toprule
                Add & Remove \\ 
                \midrule
                 $c_4$ & $\neg c_4$, $\phi(c_4)=c_1$ \\ 
                $c_5$    & $\neg c_5$, $\phi(c_5)=c_2$ \\ 
                $c_6$    & $\neg c_6$, $\phi(c_6)=c_3$ \\ 
                 $ \neg c_1$   & $c_1$ \\ 
                 $ \neg c_2$   & $c_2$ \\ 
                \bottomrule
            \end{tabular}
        };
        \node (table1) at (0,0) {
            \begin{tabular}{@{} c|c @{}}
            \toprule
                Add & Remove \\ 
                \midrule
                \rowcolor{gray!20} $c_4$ & $\neg c_4$, $c_1$ \\ 
                \rowcolor{gray!40} $c_5$    & $\neg c_5$, $c_2$ \\ 
                $c_6$    & $\neg c_6$, $c_3$ \\ 
                \rowcolor{gray!20} $ \neg c_1$   & $c_1$ \\ 
                \rowcolor{gray!40} $ \neg c_2$   & $c_2$ \\ 
                \bottomrule
            \end{tabular}
        };
        \node (table2) at (4.5,0) {
            \begin{tabular}{@{} c|c @{}}
                Add & Remove \\ 
                \midrule
                \rowcolor{gray!20} $c_4$, $\neg c_1$ & $\neg c_4$, $c_1$ \\ 
                \rowcolor{gray!40} $c_5$, $\neg c_2$    &  $\neg c_5$, $c_2$ \\ 
                $c_6$    & $c_3$, $\neg c_6$ \\                 \bottomrule
            \end{tabular}
        };
        \draw[thick, gray!40, ->] (1.35,0.6) -- (2.9,0.2);
        \draw[thick, gray!40, ->] (1.35,-0.7) -- (2.9,0.1);

        \draw[thick, gray, ->] (1.35,0.2) -- (2.9,-0.2);
        \draw[thick, gray, ->] (1.35,-1.1) -- (2.9,-0.3);
    \end{tikzpicture}
\end{center}
\caption{Illustration of the example presented in the proof of \Cref{thm:pavpositive}. The table on the left depicts the primary swaps associated with $\phi$. The right part of the figure illustrates the operation of merging two swaps.}
\label{fig:pavtables}
\end{figure}
In the general case, let $\phi$ be an arbitrary injection from $A^+\setminus W^+$ to $W^+\setminus A^+$.
Let $T\subseteq \neg D_S$ be exactly the subset of negative candidates $\neg c$ from $\neg D_S$ for which there exists some $a\in A^+\setminus W^+$ such that $\phi(a)=c$ (in our example these are candidates $\neg c_1$ and $\neg c_2$).
The collection of swaps $\mathcal{E}=\{(X_1,Y_1),\ldots, (X_r,Y_r)\}$, $r\leq d$ associated with $\phi$ (the first element of a swap corresponds to candidates to be added and the second to those to be removed)
is defined as follows:
\begin{enumerate}[itemsep=4pt]
\item Trivial swaps: If $x\in A\cap W$, $(\{x\},\{x\})\in \mathcal{E}$.
\item If $\neg x \in (D_S\setminus W)\setminus T$, then $\left(\{\neg x\},\{x\}\right)\in \mathcal{E}$.
\item If for $x \in A^+ \setminus W^+$, $\neg \phi(x)\in T$, $\left(\{x,\neg \phi(x)\},\{\neg x, \phi(x)\}\right)\in \mathcal{E}$.
\item If for $x \in A^+ \setminus W^+$, $\neg \phi(x)\notin T$, $\left(\{x\}, \{\neg {x},\phi(x)\}\right)\in \mathcal{E}$.
\end{enumerate}

Let $w_i=|U_i\cap W|$. Further, let $p$ be the number of swaps $(X,Y)\in \mathcal{E}$ such that $|X|=2$ (denote the set of these swaps as $\mathcal{E}_2$) and let $x$ be the number of such swaps where $|X|=1$ (with the set of these swaps denoted as $\mathcal{E}_1$).
Then, clearly, the following holds: 
\begin{align}
    \label{2p+x+1}
2p+x=|A|=d
\end{align}
Note that for swaps with $|X|=2$, it holds that $Y\cap A = \emptyset$, i.e., no voter approves the removed candidates. Let $\Delta_i(X,Y)$ denote the change of the PAV score that voter $i$ assigns to committee $W$ due to performing the swap $(X,Y)$. We have that:
\begin{align*}
&\sum_{(X,Y)\in \mathcal{E} } \sum_{i\in S}\Delta_i(X,Y)
= \sum_{(X,Y)\in \mathcal{E}_2}\sum_{i\in S}\left(\frac{1}{w_i+1}+\frac{1}{w_{i}+2}\right)+\sum_{(X,Y)\in \mathcal{E}_1}\left(\sum_{i\in S} \frac{1}{w_i+1}-\sum_{\substack{i\in S, \\|U_i\cap Y|=1}}\frac{1}{w_i+1}\right)\\
&=\sum_{i\in S }  \left(\frac{p}{w_i+1}+\frac{p}{w_i+2} \right)+ \sum_{i\in S} \frac{x}{w_i+1} -\sum_{i\in S} \sum_{\substack{(X,Y)\in \mathcal{E}_1, \\|U_i\cap Y|=1}} \frac{1}{w_i+1}
\geq \sum_{i\in S }  \left( \frac{p+x}{w_i+1}+\frac{p}{w_i+2} \right) -\sum_{i\in S}\frac{w_i}{w_i+1}\\
&=  \sum_{i\in S } \left( \frac{p+x}{w_i+1}+\frac{p}{w_i+2} \right) +\sum_{i\in S}\frac{1}{w_i+1} -|S| 
\geq \frac{(2p+x+1)}{2} \cdot \sum_{i\in S}\left(\frac{1}{w_i+1
} + \frac{1}{w_i+2} \right)-|S|.
\end{align*}
Now consider the swaps $\mathcal{E}(\phi_j)$ we obtain from each $\phi_j$ by the above-mentioned construction.
By summing up over all swaps that arise by all possible defined injections,
we obtain:
\begin{align}
\label{eq:positive}
\sum_{\phi_j}\sum_{(X,Y)\in \mathcal{E}(\phi_j)}\sum_{i\in S}\Delta_i(X,Y)\geq k' \cdot \left(\frac{(2p+x+1)}{2} \cdot \sum_{i\in S}\left(\frac{1}{w_i+1
} + \frac{1}{w_i+2} \right)-|S|\right) \text{.}
\end{align}

Our goal is to lower bound the following term, being the lower bound on the change in the PAV score of the voters from $N \setminus S$ due to performing the swaps: 
\begin{align*}
\Delta_{N\setminus S} = -&\sum_{\phi_j}\sum_{(X,Y)\in \mathcal{E}(\phi_j)} \left(\sum_{\substack{i\in N\setminus S,\\ |U_i\cap Y|=2}}\left( \frac{1}{w_i}+\frac{1}{w_i-1}\right)+\sum_{\substack{i\in N\setminus S,\\ |U_i\cap Y|=1}}\frac{1}{w_i}\right) \text{.}
\end{align*}
We also note that we may assume without loss of generality that if some $i\in N\setminus S$ approves $\neg c \in W^-$ then $c\in A^+$ or $\neg c \in D_S$.
Indeed, if some $i\in N\setminus S$ approves $\neg c \in W^-$ for which neither $c\in A^+$ nor $\neg c \in D_S$ holds, we may simply delete such an approval. This only results in decreasing the above term and so lower bounds the one for the original instance. This observation will become useful later on.

The proof of the following estimation lemma is rather technical, so we defer it to \Cref{app:delproofs}. As an indication that the bound is not far from being tight, we note that the proof uses inequalities only twice. Specifically, first we use the approximation that $-p_i\geq -(w_i-r_i)$ and later we also use that $\frac{1}{w_i-1}>\frac{1}{w_i}$ for $w_i>1$.
\begin{restatable}{lemma}{bound}
\label{claim:bound}
It holds that $\Delta_{N\setminus S} > -(k'+1)(n-|S|).$
\end{restatable}

We now combine \Cref{eq:positive} with the result from \Cref{claim:bound}, taking into account that $W$ is the outcome selected by PAV, and we get the following:
\begin{align*}
k'\frac{(2p+x+1)}{2} \cdot &\sum_{i\in S}\left(\frac{1}{w_i+1
} + \frac{1}{w_i+2} \right)-k'|S|-(k'+1)(n-|S|)\leq 0
\xLeftrightarrow[]{\eqref{2p+x+1}}  \notag\\ 
k' \cdot \frac{(d+1)}{2} \cdot &\sum_{i\in S}\left(\frac{1}{w_i+1
} + \frac{1}{w_i+2} \right) -k'n- n + |S| \leq 0 \Leftrightarrow \notag\\
0 \geq \frac{(d+1)}{2} \cdot &\sum_{i\in S}\left(\frac{1}{w_i+1
} + \frac{1}{w_i+2} \right) - n - \frac{n - |S|}{k'}.
\end{align*}
From that, the following follows:
\begin{align}
\label{eq:k'd+1}
0 \geq (d+1) \cdot &\sum_{i\in S}\frac{1}{w_i+\nicefrac{3}{2}
}  - n - \frac{n - |S|}{k'} \text{.}
\end{align}

From $A^+$ we did not select at most $k'$ candidates. Also, $D_S$ contains at most $k'$ negative candidates that have a positive counterpart selected. All other candidates from $D_S$ were selected. We know that the average satisfaction of the voters from $S$ equals at least $d - 2k'$. If $d - 2k' \geq d \cdot \frac{s}{n}$, then the proof follows immediately. Thus, from now on we can assume that $d - 2k' < d \cdot \frac{s}{n}$, hence:
\begin{align*}
k' > \frac{1}{2} d\cdot \frac{n-|S|}{n} \xLeftrightarrow{\eqref{eq:k'd+1}}
0 &\geq (d+1)\sum_{i\in S}\frac{1}{w_i+\nicefrac{3}{2}} - n - \frac{2n}{d} \text{.}
\end{align*}
From the inequality between harmonic and arithmetic mean we get that:
\begin{align*}
\sum_{i\in S}\frac{1}{w_i+\nicefrac{3}{2}} \geq \frac{|S|}{\avgsat_S(W) +\nicefrac{3}{2}},
\end{align*}
and combining the last two relations we get the following:
\begin{gather*}
 \frac{|S|}{\avgsat_S(W)+\nicefrac{3}{2}} < \frac{2n + nd}{d(d+1)} \Leftrightarrow
\avgsat_S(W) > \frac{d+1}{d+2}  \cdot d \frac{|S|}{n} - \frac{3}{2}\text{.} 
\end{gather*}
If $d \geq \nicefrac{k}{2}$, then the proof is complete. 
Now, we consider the case when $d < \nicefrac{k}{2}$. 
In this case however, 
we know that 
$k' = k-a \geq 
k-d >
\nicefrac{k}{2}$. 
From \Cref{eq:k'd+1} we get that:
\begin{align*}
0 \geq (d+1) \cdot &\sum_{i\in S}\frac{1}{w_i+\nicefrac{3}{2}
}  - n \cdot \frac{k'+1}{k'} \Leftrightarrow \sum_{i\in S}\frac{1}{w_i+\nicefrac{3}{2}
} \leq n \cdot \frac{k'+1}{k'(d+1)}\text{.}
\end{align*}
By using again the inequality between harmonic and arithmetic mean we get: 

\resizebox{0.96\linewidth}{!}{%
  \parbox{\linewidth}{%
\begin{align*}
\frac{|S|}{\avgsat_S(W)+\nicefrac{3}{2}} \leq n \cdot \frac{k'+1}{k'(d+1)} \Leftrightarrow \avgsat_S(W) &\geq \frac{|S|}{n} \cdot (d + 1) \cdot \frac{k'}{k' + 1} - \frac{3}{2}
\geq \frac{|S|}{n} \cdot d \cdot \frac{k' + 1}{k' + 2} - \frac{3}{2} \text{,}
\end{align*}
}%
}
where the last inequality follows from $d < \nicefrac{k}{2}$. In all cases we got:
\begin{align*}
\avgsat_S(W) \geq \frac{|S|}{n} \cdot d \cdot \frac{\nicefrac{k}{2} + 1}{\nicefrac{k}{2} + 2} - \frac{3}{2} = \frac{|S|}{n} \cdot d \cdot \left(1 - \frac{2}{k+4}\right) - \frac{3}{2} \text{.}
\end{align*}
This completes the proof.
\end{proof}

In the standard committee election setting, it is known that PAV guarantees that a group $S$ with $|A_S|\geq \ell$ and $|S|\geq \frac{\ell}{k} n$ has $avgsat_S(W)>\ell-1$, and that this is asymptotically tight \cite{AEHLSS18}.
In our setting, we can show that the additive constant of $\nicefrac{3}{2}$ is necessary.
\begin{theorem}
\label{thm:pav-lower-bound}
For every \(\eta>0\), there exists an election \(E\), an \(\ell\)-cohesive group \(S\), and a PAV outcome \(W\) such that
$    \avgsat_S(W)<\ell-\nicefrac{3}{2}+\eta.$

\end{theorem}\begin{proof}

Fix positive integers \(k\) and \(y\) with $k=y^2$. We construct an election with three disjoint sets of candidates \(A=\{a_1,\ldots,a_k\}\), \(B=\{b_1,\ldots,b_k\}\), and \(C=\{c_1,\ldots,c_y\}\). An \(\alpha\)-fraction of the voters, denoted by \(S\), approve all candidates in \(A\) and disapprove all candidates in \(B\cup C\). The remaining \((1-\alpha)\)-fraction of the voters approve all candidates in \(B\) and are neutral towards all other candidates.
Choose \(\alpha\) such that:
\[
    \frac{\alpha}{y+1}+\frac{\alpha}{y+2}<\frac{1-\alpha}{k}
\Leftrightarrow \alpha<\frac{(y+1)(y+2)}{(y+1)(y+2)+k(2y+3)}.\tag{1}
\label{eq:pav-lb-condition}
\]

We claim that the unique PAV outcome selects the positive candidates in \(B\). As usual, we identify an outcome with its positive part, assuming that the negative counterpart of every non-selected positive candidate is included. Under the outcome \(B\), each voter in \(S\) obtains satisfaction \(y\) from non-electing the candidates in \(C\), while each voter outside \(S\) obtains satisfaction \(k\).

Consider replacing one selected candidate \(b_i\in B\) by some \(a_j\in A\). This increases the satisfaction of every voter in \(S\) by \(2\): one unit from electing \(a_j\), and one unit from no longer electing the disapproved candidate \(b_i\). It decreases the satisfaction of every voter outside \(S\) by \(1\). Hence the change in the PAV score is \(\frac{\alpha}{y+1}+\frac{\alpha}{y+2}-\frac{1-\alpha}{k}\), which is negative by \Cref{eq:pav-lb-condition}. Further replacements of candidates from \(B\) by candidates from \(A\) are even less profitable, since the marginal gain for voters in \(S\) decreases while the marginal loss for voters outside \(S\) increases.

Dropping a candidate \(b_i\) without replacing it by a candidate from \(A\) is also unprofitable, since its first marginal change is \(\frac{\alpha}{y+1}-\frac{1-\alpha}{k}<0\). Subsequent such changes are again even less profitable. Finally, selecting a candidate from \(C\) is never beneficial, since such a candidate is disapproved by all voters in \(S\) and approved by no voter. Thus the unique PAV outcome selects exactly \(B\).

Now \(A_S=A\) and \(D_S=B\cup C\), so \(|A_S|=k\) and \(|D_S|=k+y\). Hence \(S\) is \(\ell\)-cohesive for \(\ell=\alpha(|D_S|+\min(k,|A_S|))=\alpha(2k+y)\). However, under the PAV outcome, the average satisfaction of voters in \(S\) is only \(y\). Thus the gap between the claim of \(S\) and its obtained satisfaction is \(\ell-y=\alpha(2k+y)-y\).

The constraint in \Cref{eq:pav-lb-condition} allows us to choose \(\alpha\) arbitrarily close to \(\alpha^\star:=\frac{(y+1)(y+2)}{(y+1)(y+2)+k(2y+3)}\). For this value, the corresponding gap is
\[
    \alpha^\star(2k+y)-y
    =
    \frac{k(3y+4)}{(y+1)(y+2)+k(2y+3)}.
\]
Taking \(k=y^2\) and letting \(y\to\infty\), this expression converges to \(\nicefrac{3}{2}\). Therefore, for every \(\eta>0\), we can choose \(y\), set \(k=y^2\), and take a rational \(\alpha<\alpha^\star\) sufficiently close to \(\alpha^\star\) so that \(\ell-y>\nicefrac{3}{2}-\eta\). Equivalently, \(\avgsat_S(B)=y<\ell-\nicefrac{3}{2}+\eta\), as required.
\end{proof}

While computing the outcome of PAV is NP-hard \citep{azi-gas-gud-mac-mat-wal:c:multiwinner-approval},
there exists a polynomial-time variant, namely local search PAV, that achieves strong proportionality guarantees in the setting without negative votes
\citep{AEHLSS18,kraiczy2024lower}. 
While local search PAV is not applicable to our setting, our proof of \Cref{thm:pav-negative-votes} suggests an alternative
(swap-based) local search procedure: using the swaps that arise in our construction of the proof, it is possible to define a process, mirroring the known local search PAV rule, that also runs in polynomial time retaining the guarantees that we established for PAV.

\subsection{When Do Vetoes Matter?}
\label{sec:counterfactual}
The guarantees above show that the symmetric model admits strong proportionality results. 
Nevertheless, because voters get satisfaction both from electing approved candidates and from excluding disapproved candidates, a group may use voting power to block candidates that would not have been elected even without its opposition. 
To see this, consider the following simple instance.

\begin{example}
\label{ex:wasted-veto}
Let \(C=\{a,b,c\}\) and \(k=1\). Three voters approve \(b\) and disapprove \(a\), while two voters approve \(c\). Under Phragm\'en's rule, the first group can afford to veto \(a\) at time \(1/3\), so \(\neg a\) is selected and their budgets are reset. The second group can then afford to select \(c\) at time \(1/2\), and since \(k=1\), the rule stops with outcome \(\{c\}\).
This veto is counterfactually irrelevant: candidate \(a\) has no supporters and would not be selected even if no one vetoed it. Indeed, if vetoes against \(a\) are deleted, then the first group instead selects \(b\) already at time \(1/3\), before the second group can select \(c\). Thus, spending budget on the veto against \(a\) prevents the first group from obtaining the candidate \(b\) that it approves, even though \(a\) had no chance of being elected.
\hfill $\lrcorner$
\end{example}
The example illustrates that a coalition may ``spend influence'' on excluding a candidate that would have not been selected in the first place, while at the same time reducing the coalition's ability to secure candidates it approves.
Moreover, the issue is not rule-specific and the same profile gives the analogous phenomenon for PAV.

Of course, even in approval voting an individual approval may not be pivotal and a candidate might have been selected even without a particular voter approving her. However, with an at-most-\(k\) committee-size constraint, if all voters withdraw their approvals from a candidate, then there is typically no reason for an approval-based rule to select that candidate. In the symmetric model, by contrast, even a group's common vetoes may be counterfactually irrelevant. 

Thus, the issue is not that the model fails to provide proportionality, but that proportionality under this interpretation may count influence that does not affect the outcome. 
The purpose of this section is to make this limitation precise and to study modifications that preserve the proportionality guarantees while avoiding such wasted veto influence.
This raises the following question:

\vspace{0.3\baselineskip}
\centerline{\textit{When are vetoes \emph{outcome-relevant}, i.e., when does a candidate have a \emph{chance of winning}?}}
\vspace{0.3\baselineskip}

To isolate which vetoes are genuinely outcome-relevant, in this section we propose a solution that relies on selecting a subset $T$ of candidates whose vetoes are ``kept''.
Fix an election $E=(N,C,(A_i)_{i\in N},(D_i)_{i\in N},k)$, where each voter $i$ approves
$A_i\subseteq C$ and disapproves $D_i\subseteq C$. For $T\subseteq C$, let $E_T$ denote the
instance obtained from $E$ by restricting disapprovals to $T$, i.e., by replacing each $D_i$
with $D_i\cap T$. Fix a resolute rule $R$.
The main idea is now to infer the proper set $T$ on behalf of voters and return the outcome that would be selected when vetoes are restricted to $T$. 
First, the vetoes should be ignored only if the corresponding candidates are not to be selected anyway. 
The following axiom ensures this.

\begin{definition}Given an election $E$, we say that a rule $R$ satisfies \emph{Counterfactual Veto Deletion} (CVD) at $T$ if for any candidate $c \notin T$, $c$ is not selected under $R$ on $E_T$.\hfill $\lrcorner$
\end{definition}

The CVD part alone already aligns naturally with group-based proportionality, as the next result shows. 
The theorem shows that although deleting vetoes may reduce a group's formal cohesiveness claim in \(E_T\), CVD ensures that the deleted commonly disapproved candidates are not selected. When the outcome is evaluated in the original election, these candidates therefore contribute exactly the satisfaction needed to offset the possible loss in the lower bound. Thus, the worst-case proportionality guarantee obtained from \(R(E_T)\) is weakly stronger than the one for \(R\) on~\(E\).
Let \(g:\naturals\to\mathbb{R}\) be nondecreasing and such that \(g(\ell)-g(\ell')\leq \ell-\ell'\) for all \(\ell'\leq \ell\).

\begin{restatable}{theorem}{pres} \label{theorem:counterfactual-preserves-guarantee}
    Suppose that \(\mathcal R\) guarantees average satisfaction at least \(g(\ell)\) to every \(\ell\)-cohesive group. Fix an election \(E\) and a set \(T\subseteq C\), and let \(W=\mathcal R(E_T)\). If \(\mathcal R\) satisfies CVD at \(T\), then for every \(\ell\)-cohesive group \(S\) in \(E\) it holds that 
\(\avgsat_S(W)\geq g(\ell).\)
\end{restatable}

In addition to CVD, vetoes should be $c$ are justified only when they are
pivotal for excluding $c$. Maximal Veto Deletion is meant to rule precisely such wasted veto effort. 
\begin{definition}Given an election $E$, we say that a rule $R$ satisfies \emph{Maximal Veto Deletion} (MVD) at $T$ if for all $c \in T\setminus R(E_T)$ it holds that $c$ is winning in $E_{T\setminus\{c\}}$.\hfill $\lrcorner$
\end{definition}
If MVD holds at \(T\), then no further veto can be deleted while maintaining the counterfactual justification. Thus MVD serves as a maximality condition for the vetoes that are removed.

Note that every rule trivially satisfies CVD at $T=C$ and satisfies MVD at $T=
\emptyset$. Thus, we seek rules that satisfy both at the same set. We say that $R$ is \emph{counterfactually stable} if for every election $E$ there exists $T\subseteq C$
such that $R(E_T)$ satisfies CVD at $T$ and MVD at $T$. In this case we also say that $R$ is counterfactually stable \emph{at} $T$. While counterfactual stability is desirable as a justification for deleting vetoes on behalf of voters, we show that unfortunately, the rules examined in \Cref{bpjrphragmen,thm:pav-negative-votes} need not be counterfactually stable. In fact, this remains true even if CVD is replaced by the weaker requirement that a deleted-veto candidate \(c\notin T\) may be selected under \(R(E_T)\), provided that \(c\) would still be selected under \(R(E_{T\cup\{c\}})\), which we discuss in \Cref{app:counterfactual}.

\begin{restatable}{proposition}{phragmenfailscs}
\label{prop:phragmen-fails-cs}
There exists an election \(E\) such that, for every \(T\subseteq C\), neither PAV nor Phragm\'en's rule is counterfactually stable at \(T\).
\end{restatable}

Since Phragm\'en's rule and PAV need not admit a veto set at which they satisfy both CVD and MVD, we consider two weaker approaches. The first is a history-dependent refinement designed to avoid vetoes that are locally irrelevant to excluding their target. The second retains the static CVD framework but replaces MVD with an inclusion-maximality condition that is always achievab

Since neither Phragm\'en nor PAV are counterfactually stable in general, we retain CVD---which is sufficient for preserving our proportionality guarantees---and consider two relaxations of MVD.

The first approach is to greedily delete vetoes while preserving CVD until the resulting veto set is inclusion-maximal. This yields the weaker Inclusion Maximal Veto Deletion (IMVD) condition that is always achievable. 
The second approach is a sequential refinement of MVD. It constructs a CVD veto set sequentially and ensures that every veto that is actually executed is necessary to prevent its target from being selected, given the state of the execution at the moment the veto is considered. Thus, both approaches produce a veto set satisfying CVD, but they provide different alternatives to MVD.

\medskip

We say that a rule $R$
satisfies \emph{Inclusion-Maximal Veto Deletion} (IMVD) at $T$ if, for every candidate $c\in T$,
whenever $c$ is \emph{not} selected under $E_{T}$, removing $c$ from $T$ would create a
CVD violation; that is, $R$ fails CVD at $T\setminus\{c\}$. Intuitively, IMVD means that no further
veto can be deleted while preserving CVD, which enables a greedy construction of $T$.
IMVD is strictly weaker than MVD---it is weaker by definition and strictness is witnessed by the instance used in the proof of \Cref{prop:phragmen-fails-cs}.

\begin{proposition}
For every rule $R$ and every election $E$, there exists a set $T\subseteq C$ such that $R$ satisfies CVD and IMVD at $T$. Moreover, such a set $T$ can be computed by a greedy algorithm in polynomial time, provided that $R$ is polynomial-time computable.
\end{proposition}

The second relaxation exploits the sequential nature of Phragm\'en's rule. Whenever a veto against a candidate $c$ becomes affordable, we consider the continuation in which vetoes against $c$ are removed. The veto is executed only if $c$ is selected in this continuation. The resulting veto set satisfies CVD, while every executed veto is pivotal at the point at which it is used, though not necessarily in the global sense required by MVD.
\medskip

\begin{mdframed}[linecolor=black, linewidth=0.3mm, innerleftmargin=10pt, leftmargin=10pt, rightmargin=10pt] \textbf{Sequential Counterfactual Phragm\'en.} Fix a resolute tie-breaking rule. Given an election \(E\), maintain an \textit{active veto set} \(T\subseteq C\), initially \(T\gets C\), and run Phragm\'en's rule for positive and negative votes as defined in \Cref{sec:phragmen}, while allowing vetoes only against candidates in \(T\). Whenever some \(c\in T\) is \textit{about to be vetoed}, perform a counterfactual check: recursively simulate the continuation from the current state with vetoes against \(c\) deleted, i.e., with active set \(T\setminus\{c\}\), using the same procedure and tie-breaking for all later veto events. If \(c\) is not selected in this recursive continuation, set \(T\gets T\setminus\{c\}\) and do not veto \(c\). Otherwise, keep \(c\) in \(T\), execute the veto, and charge its vetoers as in Phragm\'en's rule. The rule returns the selected positive candidates at termination. \end{mdframed}
\medskip
Although we focused on Phragm\'en's rule for ease of presentation, the same idea extends to any rule $R$; see \Cref{app:counterfactual}.

\section{The Asymmetric Utility Model}
\label{sec:asym}
The utility model examined in \Cref{sec:uniform} enforces symmetry, treating voters as equally concerned with electing approved candidates and preventing the selection of disapproved ones. However, this symmetry may not always reflect voters' true preferences. In some scenarios, voters might rightfully feel entitled to a certain number of representatives in the elected body, regardless of whether their negative votes contributed to blocking certain candidates.
To account for this, we introduce the \emph{asymmetric utility model}, where voters' satisfaction depends only on the election of approved candidates, while disapprovals serve to hinder a candidate’s selection.
This imbalance requires introducing separate guarantees with respect to (i) approved, and (ii) disapproved candidates.

\subsection{Axioms for Positive Representation and the Group Veto Axiom}
\label{sec:nonuniform axioms}

In this section, we propose two axioms which formalize voters' entitlement to (i) representation in the committee and (ii) the ability to veto candidates.
Our first axiom ensures guarantees for voters with shared \emph{positive preferences}. It is inspired by the classic Extended Justified Representation (EJR) for approval-based committee elections and reduces to it when $D_i = \emptyset$ for all $i \in V.$
We propose a definition of cohesiveness in the examined setting which defines how large a group $S$ must be, in the presence of voters with opposing preferences, to justify electing $\ell$ representatives of their liking. 
Specifically, we say that $S$
must have enough voters to outweigh the effect of vetoing voters—meaning that for each negative vote, there must be a voter in $S$ to cancel it out—and still remain large enough to claim a fair share.

\begin{definition}[\textbf{$\ell$-positive-cohesiveness}]
Consider an election $E$. 
Given a positive integer $\ell \leq k,$ we say that a set of voters $S\subseteq V$ is $\ell$-positively-cohesive if there exists a set of at least $\ell$ candidates $T$ such that $T\subseteq A_S$ and $|S|-|D_c|\geq \ell\cdot\nicefrac{n}{k}$ for every candidate $c\in T$.\hfill $\lrcorner$
\end{definition}
After defining $\ell$-positive-cohesivess, we can provide the analogs of EJR and PJR in this setting. We say that an outcome $W$ for an election $E$ provides \textbf{Extended Justified Positive Representation (EJPR)} if for each $\ell$-positively-cohesive group of voters $S$ there exists a voter $i\in S$ such that $|A_i\cap W| \geq \ell$, and \textbf{Proportional Justified Positive Representation (PJPR)} if for each $\ell$-positively-cohesive group of voters $S$, it holds that $|\cup_{i\in S}A_i\cap W| \geq \ell$.
 A rule $\mathcal{R}$ satisfies EJPR (resp. PJPR) if for every election $E$ its winning outcome provides EJPR (resp. PJPR).

 \medskip

Having defined a guarantee for commonly approved candidates, we now aim to define a \emph{negative guarantee} as well. This will ensure that sufficiently large groups of voters also have a right to block candidates they dislike from being selected. The necessity of such an axiom is evident as EJPR can be easily satisfied by disregarding negative votes---hence dissolving any hopes for negative guarantees---and then applying any rule that satisfies EJR in the approval-based committee elections setting. On the flip side, the na\"ive rule that always returns the empty committee fully respects negative votes, but offers no positive guarantees.
Thus, from a rule-design standpoint, the goal is to develop a rule that offers both positive and negative guarantees simultaneously.
The axiom we propose, called \textit{Group Veto}, aims to ensure the exclusion from the committee of candidates based on the vetoes they receive from sufficiently large groups of voters. 
In other words, a strong opposition limits the number of candidates that can be elected from the opposed set.

\begin{definition}[\textbf{Group Veto}]
Consider an election $E$. 
For a set of candidates $T,$ say that 
$ap(T)$ is the set of voters that approve at least one candidate from $T$.
Given a non-negative integer $\ell \leq k$ and a set of at least $\ell$ candidates $T$, we say that a set of voters $S\subseteq V$ is $(\ell,T)$-negatively-cohesive if $T\subseteq D_S$ and $|S|\geq |ap(T)|- \ell\cdot\nicefrac{n}{k}.$ An outcome $W$ is said to provide Group Veto for $E$ if for each $(\ell,T)$-negatively-cohesive group it holds $|W \cap T|\leq \ell$. A rule $\mathcal{R}$ satisfies the Group Veto axiom if for every election $E$ its winning outcome provides Group Veto.\hfill $\lrcorner$
\end{definition}

\noindent 
 \textbf{Continuation of \Cref{ex:intro_example}.} Since for both groups of voters $V_1, V_2$, we can find approved sets (respectively) $C_3, C_1$ that are not opposed by the other group, the guarantees given by both EJPR and PJPR are $6$ for $V_1$ and $4$ for $V_2$. Note that by themselves they do not say whether the guarantee for $V_1$ should be satisfied by candidates from $C_2$ or from $C_3$. 
  Setting $S=V_2$ and $T = C_2$ in the Group Veto definition, we find that $\ell = (|ap(T)| - |S|)\cdot \nicefrac{k}{n} = (|V_1| - |V_2|)\cdot \nicefrac{k}{n} = 2$. As a result, 
any rule satisfying the Group Veto axiom should
elect at most 2 candidates from $C_2$. 
\hfill $\lrcorner$

\smallskip
We propose rules that fit to the interpretation of utilities in the asymmetric model and we analyze them in terms of whether they satisfy the proposed axioms simultaneously.

\subsection{Method of Equal Shares and Phragm\'en's Rule with Opposition Tax}

According to the original definitions of the Method of Equal Shares and Phragm\'en's rule (\Cref{sec:prelims}) candidates are assumed to have unit costs.
In our generalizations, we increase them by introducing an \emph{opposition tax}. 
In turn, we call the rules \textit{Method of Equal Shares} and \textit{Phragm\'en with Opposition Tax} (\textit{Tax-MES} and \textit{Tax-Phragm\'en} for short).
The tax captures the idea that the more voters veto a candidate $c$, the higher the price her supporters need to pay to elect $c$.
As a result, candidates' prices become unequal, setting our methods apart from their standard analogs.
We select the tax appropriately to ensure (i) among candidates with equal number of supporters, the one with higher net approval is prioritized, (ii) among those with equal net approval, the one with more supporters is prioritized. We focus on electing only candidates with more supporters than opponents.

\medskip
\begin{mdframed}[linecolor=black, linewidth=0.3mm, innerleftmargin=10pt, leftmargin=10pt, rightmargin=10pt
]
\textbf{Tax-MES/Tax-Phragm\'en.}
     Given an election $E$, we define $C'=\{c\in C: |A_c|>|D_c|\}$. For each $c\in C'$ we set its price to $p(c):=\frac{|A_c|}{|A_c|-|D_c|}.$ We create a ballot profile $B'$ such that 
     for each voter $i \in V,$ we set
     $B_i':=(A_i,\emptyset), \forall i\in V.$ Then, in $(C',V,k,B')$ we execute MES (respectively, Phragm\'en) in their standard form. In the case of Phragm\'en, we stop increasing a voter’s budget once she has received $\nicefrac{k}{n}$ in total.
    \end{mdframed}

\medskip
It is straightforward to check that in the instance of \Cref{ex:intro_example} both tax-MES and tax-Phragm\'en elect outcomes that satisfy the aforementioned proportionality axioms. Specifically, since the price is $1$ for candidates in $C_1 \cup C_3$ and $3$ for $C_2$, both rules elect $6$ candidates from $C_3$ and $4$ from $C_1$. This observation can be generalized as \Cref{thm:mes} below. Note that Tax-Phragm\'en, must fail EJPR since if $D_i=\emptyset$ for all $i\in V$, then it reduces to classic Phragm\'en, which fails EJR \cite{brill2024phragmen}. 

\begin{restatable}{theorem}{tmes}\label{thm:mes}
Tax-MES satisfies EJPR and Group Veto. Tax-Phragm\'en satisfies PJPR and Group Veto.
\end{restatable}

\subsection{On PAV and (Generalized) Thiele Rules in the Asymmetric Setting}
In \Cref{thm:pav-negative-votes} we proved that PAV exhibits strong proportionality guarantees in the symmetric utility model. Interestingly, the rule is not suitable if we take the asymmetric interpretation.
In particular, below, we show that PAV cannot be adapted to satisfy EJPR while accounting for negative votes. This limitation extends to the entire family of Thiele rules when adapted for this setting: the class of \textit{generalized Thiele rules}. Each rule is defined via a mapping $f$, where $f(z,s)$ specifies the score that a committee obtains from
a voter that approves $z$ candidates in it while disapproving $s$. 

\medskip
\begin{mdframed}[linecolor=black, linewidth=0.3mm, innerleftmargin=10pt, leftmargin=10pt, rightmargin=10pt]
\textbf{Generalized Thiele Rules.}
A generalized Thiele rule induced by the scoring function $f\colon \naturals^2 \to \reals$ selects for each election $E$ a feasible outcome $W$ that maximizes: $\sum_{i\in V}f(|W\cap A_i|,|W\cap D_i|).$
\end{mdframed}

\begin{restatable}{theorem}{thmthiele}
\label{thm:Thiele}
    No generalized Thiele rule induced by a scoring function $f$ such that $f(z, s) < f(z, 0)$ for some $(z, s)\in \naturals^2$, satisfies EJPR.
\end{restatable}
\begin{proof}
Consider a generalized Thiele rule induced by a function $f$ for which the desired pair of values $(z, s)\in \naturals^2$ exists. Assume for the sake of contradiction that the rule satisfies EPJR. Since in the standard setting EJPR is equivalent to standard EJR, and PAV is the only Thiele rule satisfying EJR \cite{aziz2017justified}, it holds that $f(x, 0) = \sum_{1 \leq i \leq x} \nicefrac{1}{i}$ for every $x\in \naturals$. 
We construct a family of instances parameterized by $z$ and $s$. Specifically, for any value of $t>s,$ we define an election $E_t=(C,V,k,B)$ as follows: The set of candidates $C$ consists of $3(t+z)$ candidates in total, $2(t+z)$ of which are dummy candidates who are neither approved nor disapproved by any voter. Apart from the dummy candidates, we have 3 disjoints sets of candidates: $T_1$ of $t-s$ candidates, $T_2$ of $s$ candidates and $T_3$ of $z$ candidates. There are two disjoint groups of voters: $V_1$, of size $\nicefrac{2n}{3}$, approves $T_1 \cup T_2 \cup T_3$, and $V_2$, of size $\nicefrac{n}{3}$, approves $T_3$ and disapproves $T_2$.
We set $k= 3(t+z)$. 
EJPR requires that voters in $V_1$ deserve at least $t+z$ approved candidates in the committee, hence the considered rule should elect all candidates from $T_1 \cup T_2 \cup T_3$.
Nevertheless, using the obtained relation for $f(x, 0)$, we get that there exists a value of $t$ such that, in the election $E_t$, the committee $T_1 \cup T_3$ is better according to $f$. Such an election $E_t$ serves as the counterexample establishing the theorem.
\end{proof}
The assumption appearing in \Cref{thm:Thiele} simply says that no voter should prefer a committee containing candidates they disapprove over one without (as long as both include the same number of candidates they approve)---a natural restriction.
A direct consequence of \Cref{thm:Thiele} is that no generalized Thiele rule satisfies any form of a negative guarantee, such as Group Veto, while also satisfying EJPR. In particular, there does not exist a reasonable (i.e., not neglecting negative votes) extension of PAV to this setting. Importantly, \Cref{thm:Thiele} also applies to PJPR without any changes to the construction used in its proof.
The markedly different behavior of the two PAV variants we study in \Cref{sec:uniform} and \Cref{sec:asym} reinforces the intuition that the two interpretations of negative votes we employ necessitate for different approaches---both conceptually and technically---when designing proportional voting rules for each utility model.

\section{Conclusion}

Our two formal approaches capture fundamentally different interpretations of negative votes, and the appropriate choice between them is context-dependent. The symmetric model is most natural in settings where the goal is to aggregate voters' \emph{opinions} fairly, rather than to provide direct representation within the selected set. Examples include legislation, the selection of statements, blog posts, or comments, etc. (cf., \Cref{sec:applications}). In contrast, the asymmetric model is better suited to settings where the selected candidates themselves are meant to provide \emph{representation}, such as participatory budgeting elections or elections to representative bodies.

Taken together, our results show that moving beyond binary approval ballots gives rise to genuinely new proportionality questions, affecting both the choice of axioms and the design of voting rules. We hope that our framework will inspire further work on richer election models with approvals, disapprovals, and abstentions, beyond the concepts introduced in this paper.

\smallskip
Regarding the utility models studied here, there are several additional avenues for future work. First, it remains open to close the gap in our analysis of PAV in the symmetric model. \Cref{thm:pav-negative-votes} guarantees average satisfaction at least \((1-\varepsilon)\ell-\nicefrac{3}{2}\), while \Cref{thm:pav-lower-bound} shows that the additive constant \(\nicefrac{3}{2}\) is unavoidable. Whether the additional \(\varepsilon\ell\) loss is inherent to PAV, or merely an artifact of the proof, remains open. More broadly, it would be interesting to identify the optimal Thiele score sequence for the symmetric utility model; for instance, the shifted PAV score \(h(r)=\sum_{i=1}^{r}\nicefrac{1}{i+1}\) appears to trade a worse additive constant for the absence of the multiplicative loss in the known analysis. Second, the feasibility constraints in the symmetric model can be viewed as the intersection of two partition matroids: one imposing the cardinality bound on positive candidates, and one enforcing that at most one of \(c\) and \(\neg c\) is selected for each original candidate \(c\). This suggests studying whether our analysis extends to intersections of two partition matroids, and more generally to intersections of two matroids, or of a constant number of matroids. Third, designing efficiently computable counterfactually stable rules with non-trivial proportionality guarantees, or proving non-existence, remains an important open problem.

In the asymmetric model, our rules are currently non-exhaustive; in fact, early stopping is necessary to satisfy Group Veto. Designing proportionality guarantees that are compatible with exhaustiveness is therefore an important direction for future research. 
Another aspect of this model that deserves attention is that, under the underlying interpretation of utilities, voters incur no cost for expressing negative votes. As a result, they may vote against every project they do not approve of, regardless of whether they are actually opposed to it or simply indifferent. In some settings, this may create the need to either limit the number of negative votes a voter can cast or introduce incentives that discourage voting against projects towards which the voter is merely neutral.
We discuss these two identified issues in more detail and present initial findings in \Cref{app:exhaustive}.

%
\begin{acks}
The authors are grateful to the reviewers of earlier versions of this manuscript for their valuable feedback.
Sonja Kraiczy was supported by an EPSRC studentship 1327875. Georgios Papasotiropoulos, Grzegorz Pierczyński and Piotr Skowron have been supported by the European Union (ERC, PRO-DEMOCRATIC, 101076570).
Views and opinions expressed are however those of the author(s) only and do not necessarily
reflect those of the European Union or the European Research Council. Neither the European
Union nor the granting authority can be held responsible for them. Additionally, G. Pierczyński
was supported by Poland’s National Science Center grant no. 2022/45/N/ST6/00271.

\vspace{0.6cm}
\centering
\includegraphics[width=0.45\textwidth]{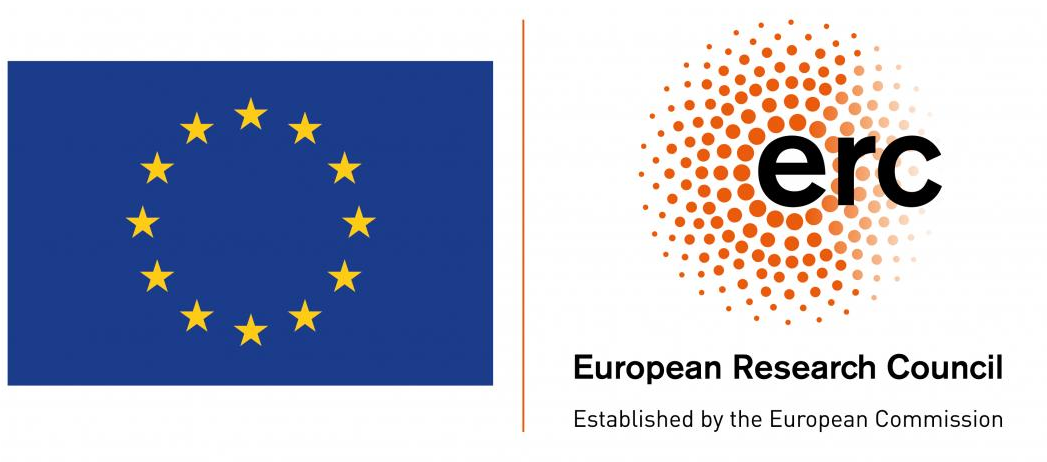}
\vspace{0.6cm}

\end{acks}

\bibliographystyle{ACM-Reference-Format}
\bibliography{sample-bibliography}

\clearpage
\appendix
{\large{\textbf{Technical Appendix}}}
\section{Supplementary Material for the Symmetric Setting}

\subsection{Omitted Proofs for Phragm\'en's Rule and PAV}
\label{app:delproofs}

\thmPJR*

\begin{proof}
We begin with Statement (1) on PJR.
Consider a group $S$ that deserves $\ell$ candidates and suppose $|\cup_{i\in S} (A_i \cup \neg D_i)\cap W|<\ell$ where $W$ is the outcome returned by the proposed rule.
Consider the first moment, $t$, when all the candidates from
$A_S \cup \neg D_S$ are either elected or removed. Note that $t < \frac{\ell}{|S|}$. Indeed, if $t \geq \frac{\ell}{|S|}$, then at time $\frac{\ell}{|S|}$
the group $S$ would collect in total $\ell$ dollars, and would buy at least $\ell$ candidates from 
$\cup_{i\in S} A_i\cup \neg D_i$ (the
possibility of buying such candidates comes from the fact that there would always be a candidate
from $A_S\cup \neg D_S$ available for purchase).
Thus, at time $t$ the number of selected candidates was lower than:
\begin{align*}
tn \leq \frac{\ell}{|S|} \cdot n = \nicefrac{|S|}{n}(|D_S|+\min(k,|A_S|) \cdot \nicefrac{n} {|S|} = |D_S|+\min(k,|A_S|) \text{.}
\end{align*}
Assume we selected at most $x$ candidates from $D_S$ and $y$ candidates from $A_S$. Since we cannot add any additional candidate from $D_S$ it means that $|D_S| - x$ candidates opposite to those from $D_S$ were selected. Clearly, none of those was approved by $S$ because approval and disapproval sets are disjoint. Since we cannot add also a candidate from $A_S$ at least one of two must hold:
\begin{enumerate}
\item at least $|A_S| - y$ candidates opposite to those from $A_S$ were selected. But then the total number of candidates selected would be at least $|A_S| + |D_S|$, a contradiction.
\item at least $k$ positive candidates were selected in total. This however means that $t \geq \nicefrac{k}{n}$, and so that the voters from $S$ paid for at least $|S| \cdot \nicefrac{k}{n}$ candidates, a contradiction.
\end{enumerate}

We now move to Statement (2) on proportionality degree. Consider an $\ell$-cohesive group $S\subset N$. 
Toward a contradiction assume that the average satisfaction that group $S$ has for $W$, the outcome returned by Phragm\'en, is less than $\frac{\ell-1}{2}$. As in the proof of Theorem 9 in the paper of \citet{masavrik2024generalised} we
define
$t=\frac{\ell}{|S|}+\frac{\Delta-1}{n} \text{,}$
where $\Delta$ is the smallest non-negative value such that at $t$ the voters from $S$ have at most $\Delta$ unspent dollars (if such $\Delta$ does not exist, then we simply set $\Delta = 0$).
Observe that $t\geq \frac{\ell}{|S|}-\frac{1}{n}\geq \frac{\ell-1}{|S|}.$
Consider the set $W$ of candidates elected by  Phragm\'en up to time $t$ and observe that it follows from Statement (1) on PJR that $W$ contains at least $\ell-1$ candidates from $\cup_{i\in S} A_i$. Similarly, we observe that at that time some candidate jointly approved by $S$ was available (as in the previous statement).
 The remaining proof continues as in \citep{masavrik2024generalised}.\looseness-1
\end{proof}
\sv{\bound*}
\begin{proof}
Let us introduce some further notation that will prove highly useful in the careful analysis that follows. For each voter $i\in N$, we define:
\begin{align*}
 b  & = |\neg D_S\cap W^-| \quad \quad
 & b_i & = |\neg D_i\cap (\neg D_S\cap W^-)| \\
 a  & = |A^+\cap W^+| \quad \quad
 & a_i & = |A_i \cap (A^+\cap W^+)| \\
 r_i & = |\neg D_i\cap W^-| \quad \quad
 & w_i & = |U_i\cap W| \\
 p_i & = |\{c \in D_S\mid c\in A_i\cap W^+\}|
\end{align*}
Note that $p_i\leq w_i-r_i-a_i$ and that $b_i = b, p_i = 0$ for $i\in S$.
Let $\mathcal{E}=\cup_{j=0}^{k'-1} \mathcal{E}(\phi_j)$.
For each agent $i$ the trivial swaps contribute $-\frac{k'(a_i+b_i)}{w_i}$ to $\Delta_{N\setminus S}$. 
Then, the counting of the number of swaps of the form $(\neg c, c)$ with $\neg c \in D_S$ and $c\in W^+\setminus A^+$ can be done as follows. Among the $k'$ injections $\phi_1,\ldots,\phi_{k'}$, whenever $c$ is in the image of a $\phi_i$, the initial exchange $(\neg c, c)$ will be merged into an exchange of the form $(\{\neg c, \phi_i^{-1}(c)\}, \{\phi_i(c), c\})$. So we need to count the number of these injections for which the exchange $(\neg c, c)$ does not get merged this way. Since each of the $k'$ injections has an image of size $d^+-a$ and each of the $k'$ candidates $c\in W^+\setminus A^+$ is in the image of the same number of such injections, each of them is in the image $\frac{k'(d^+-a)}{k'}=d^+-a$ times. Hence, the candidate $c$ is not in the image of $k'-(d^+-a) = k - d^+$ injections. Thus, the swaps of the form $(\neg c, c)$ for the voter $i\in N\setminus S$ contribute $-(k-d^+)\cdot\frac{p_i}{w_i}$ to the above negative term $\Delta_{N\setminus S}$. We have now covered all the swaps $(X,Y)\in \mathcal{E}$ for which $|Y|=1$.

We next count the swaps $(X,Y)\in \mathcal{E}$ where $|Y|=2$. These are swaps that involve removing a negation $\neg c$ and another positive candidate $g$ in $W^+ \setminus A^+$ as well as adding candidate $c\in A^+$ (and potentially adding $\neg g$).
So the set $Y$ has the form $\{\neg c, g\}$ where $g\in W^+\setminus A^+$ and $c\in A^+\setminus W^+$.

We will split these swaps further into these for which agent $i$ approves (1) only $\neg c$, (2) only $g$ or (3) both of them. (It is also possible that $i$ approves neither of them; however, then the removal of those candidates does not decrease the score the voter assigns to $W$.)

The number of candidates $g\in W^+\setminus A^+$ that $i$ approves is $w_i-r_i-a_i$.
The number of candidates $g\in W^+\setminus A^+$ that $i$ does not approve is the rest of them, i.e., $(k'-(w_i-r_i-a_i))$.
The number of candidates $c\in A^+\setminus W^+$ such that $\neg c \in W^-$ and $i$ disapproves $c$ is $(r_i-b_i)$ (where we use the fact that if $i\in N\setminus S$ disapproves $c$ then either $c\in A^+$ or $c\in D_S$).
The number of candidates $c\in A^+\setminus W^+$ such that $i$ does not disapprove $c$ is the rest of them i.e. $|A^+\setminus W^+|-(r_i-b_i)=d^+-a-(r_i-b_i)$.
Consequently, the number of pairs in which $i$ both disapproves $c$ and approves $g$ is $(r_i-b_i)(w_i-r_i-a_i)$,
the number of pairs in which $i$ disapproves $c$ but does not approve $g$ is $(r_i-b_i)(k-a-(w_i-r_i-a_i))$ and finally the number of pairs in which $i$ approves $g$ but does not disapprove $c$ is $(d^+-a-(r_i-b_i))(w_i-r_i-a_i)$.

Now, observe that the reason why $|Y| = 2$ is that for a positive candidate $c$ from $A^+ \setminus W$ we are adding we also need to remove $\neg c$ which is in the committee. In each of the $k'$ injections $c$ is matched with a different candidate from $W^+ \setminus A^+$ to form such a swap $(X,Y)$. Thus, each set $Y$ appears at most once in all the injections.
Collecting all the terms we get the following equality:
\begin{align*}
\Delta_{N\setminus S}  = &-\sum_{\phi_j}\sum_{(X,Y)\in \mathcal{E}(\phi_j)} \sum_{\substack{i\in N\setminus S,\\ |U_i\cap Y|=2}}\left( \frac{1}{w_i}+\frac{1}{w_i-1}\right)+\sum_{\substack{i\in N\setminus S,\\ |U_i\cap Y|=1}}\frac{1}{w_i}\\
\geq &-\sum_{i\in N\setminus S} (r_i-b_i)(w_i-r_i-a_i) \left(\frac{1}{w_i}+\frac{1}{w_i-1}\right) -\sum_{i\in N\setminus S} ((d^+-a)-(r_i-b_i))(w_i-r_i-a_i)\frac{1}{w_i}\\
&-\sum_{i\in N\setminus S}(r_i-b_i)(k'-(w_i-r_i-a_i))\frac{1}{w_i} -\sum_{i\in N\setminus S} k'(a_i+b_i)\frac{1}{w_i}-\sum_{i\in N\setminus S} (k-d^+)p_i \frac{1}{w_i}.
\end{align*}
We now proceed to simplify the summations. 
\begin{align*}
\Delta_{N\setminus S}  =& -\sum_{\phi_j}\sum_{(X,Y)\in \mathcal{E}_j} \sum_{\substack{i\in N\setminus S,\\ |U_i\cap Y|=2}}\left( \frac{1}{w_i}+\frac{1}{w_i-1}\right)+\sum_{\substack{i\in N\setminus S,\\ |U_i\cap Y|=1}}\frac{1}{w_i}\\
\geq  &-\sum_{i\in N\setminus S} (r_i-b_i)(w_i-r_i-a_i) \left(\frac{1}{w_i}+\frac{1}{w_i-1}\right)\\
&-\sum_{i\in N\setminus S} (\textcolor{RoyalBlue}{(d^+-a)}-(r_i-b_i))\textcolor{RoyalBlue}{(w_i-r_i-a_i)}\frac{1}{w_i}\\
&-\sum_{i\in N\setminus S}(r_i-b_i)(k'-(w_i-r_i-a_i))\frac{1}{w_i}\\
& -\sum_{i\in N\setminus S} k'(a_i+b_i)\frac{1}{w_i}-\sum_{i\in N\setminus S} (k'-\textcolor{RoyalBlue}{(d^+-a)})\textcolor{RoyalBlue}{p_i} \frac{1}{w_i}
\end{align*}
Adding up the highlighted terms and using the fact that $p_i\leq w_i-r_i-a_i,$ the above is at least:
\begin{align*}
& -\sum_{i\in N\setminus S} \textcolor{RoyalBlue}{(r_i-b_i)(w_i-r_i-a_i) \left(\frac{1}{w_i}+\frac{1}{w_i-1}\right)}\\
&-\sum_{i\in N\setminus S} (-(r_i-b_i))(w_i-r_i-a_i)\frac{1}{w_i}\\
&-\sum_{i\in N\setminus S}(r_i-b_i)(k'-(w_i-r_i-a_i))\frac{1}{w_i}\\
& -\sum_{i\in N\setminus S} k'(a_i+b_i)\frac{1}{w_i}-\sum_{i\in N\setminus S} k'(w_i-r_i-a_i) \frac{1}{w_i}\\
=& -\sum_{i\in N\setminus S} (\textcolor{RoyalBlue}{r_i-b_i})\textcolor{RoyalBlue}{(w_i-r_i-a_i)} \frac{1}{w_i}\\
&-\sum_{i\in N\setminus S} (r_i-b_i)(w_i-r_i-a_i) \frac{1}{w_i-1}\\
&-\sum_{i\in N\setminus S} (\textcolor{RoyalBlue}{-(r_i-b_i)})\textcolor{RoyalBlue}{(w_i-r_i-a_i)}\frac{1}{w_i}\\
&-\sum_{i\in N\setminus S}(r_i-b_i)(k'-(w_i-r_i-a_i))\frac{1}{w_i}\\
& -\sum_{i\in N\setminus S} k'(a_i+b_i)\frac{1}{w_i}-\sum_{i\in N\setminus S} k'(w_i-r_i-a_i) \frac{1}{w_i}.
\end{align*}
In the last step we simply split up the first term and obtained further cancellations.
Next, we simplify the $\frac{1}{w_i-1}$ term and also obtain further cancellations.
\begin{align*}
\geq&-\sum_{i\in N\setminus S} \textcolor{RoyalBlue}{(r_i-b_i)(w_i-r_i-a_i) \left(\frac{1}{w_i-1}\right)}\\
&-\sum_{i\in N\setminus S}(r_i-b_i)(k'-(w_i-r_i-a_i))\frac{1}{w_i}\\
& -\sum_{i\in N\setminus S} k'(a_i+b_i)\frac{1}{w_i}-\sum_{i\in N\setminus S} k'(w_i-r_i-a_i) \frac{1}{w_i}\\
=&-\sum_{i\in N\setminus S} (r_i-b_i)-\textcolor{RoyalBlue}{\sum_{i\in N\setminus S}- \left(\frac{(r_i-b_i)(r_i+a_i-1)}{w_i-1}\right)}\\
&\textcolor{RoyalBlue}{-\sum_{i\in N\setminus S}(r_i-b_i)(k'-(w_i-r_i-a_i))\frac{1}{w_i}}\\
& -\sum_{i\in N\setminus S} k'b_i\frac{1}{w_i}-\sum_{i\in N\setminus S} k'(w_i-r_i) \frac{1}{w_i}
\end{align*}

Finally, it remains to iteratively collect terms that cancel.
\begin{align*}
>
&-\sum_{i\in N\setminus S} (r_i-b_i)-\sum_{i\in N\setminus S}- \left(\frac{(r_i-b_i)(r_i+a_i-1)}{w_i}\right)\\
&-\sum_{i\in N\setminus S}\frac{\textcolor{RoyalBlue}{(r_i-b_i)k'}}{w_i}-\sum_{i\in N\setminus S}-\frac{(r_i-b_i)(w_i-r_i)}{w_i}-\sum_{i\in N\setminus S}-\frac{(r_i-b_i)(-a_i)}{w_i}\\
& -\sum_{i\in N\setminus S} \textcolor{RoyalBlue}{k'b_i}\frac{1}{w_i}-\sum_{i\in N\setminus S} \textcolor{RoyalBlue}{k'}(w_i\textcolor{RoyalBlue}{-r_i}) \frac{1}{w_i}\\
= 
&-\sum_{i\in N\setminus S} \textcolor{RoyalBlue}{(r_i-b_i)}-\sum_{i\in N\setminus S}- \left(\frac{(r_i-b_i)(r_i+a_i-1)}{w_i}\right)\\
&-\sum_{i\in N\setminus S}-\frac{\textcolor{RoyalBlue}{(r_i-b_i)}(\textcolor{RoyalBlue}{w_i}-r_i)}{w_i}-\sum_{i\in N\setminus S}-\frac{(r_i-b_i)(-a_i)}{w_i}\\
& -\sum_{i\in N\setminus S} kw_i \frac{1}{w_i}\\
=
&-\sum_{i\in N\setminus S}- \left(\frac{\textcolor{RoyalBlue}{(r_i-b_i)}(\textcolor{RoyalBlue}{r_i}+a_i-1)}{w_i}\right)\\
&-\sum_{i\in N\setminus S}\textcolor{RoyalBlue}{-\frac{(r_i-b_i)(-r_i)}{w_i}}-\sum_{i\in N\setminus S}-\frac{(r_i-b_i)(-a_i)}{w_i}\\
& -\sum_{i\in N\setminus S} k' w_i \frac{1}{w_i}\\
= 
&-\sum_{i\in N\setminus S}- \left(\frac{(r_i-b_i)(a_i-1)}{w_i}\right)
-\sum_{i\in N\setminus S}-\frac{(r_i-b_i)(-a_i)}{w_i}
 -\sum_{i\in N\setminus S} k'w_i \frac{1}{w_i}\\
 &= -\sum_{i\in N\setminus S} \left(k' +\frac{r_i-b_i}{w_i} \right) \geq -\sum_{i\in N\setminus S} \left(k' +\frac{r_i}{w_i} \right) 
 \geq -\sum_{i\in N\setminus S} \left(k' + 1 \right) = -(n-|S|)(k' + 1)\text{.} 
\end{align*}Therefore, we obtained the required expression.\qedhere
\end{proof}

\subsection{Omitted Proofs and Further Details on Counterfactual Stability and Related Ideas}\label{app:counterfactual}
\pres*

\begin{proof}
Fix an \(\ell\)-cohesive group \(S\) in the original election \(E\), and let
\(W=\mathcal R(E_T)\). Let \(R=D_S\setminus T\) be the set of candidates commonly disapproved by \(S\) whose vetoes are removed in \(E_T\). In the modified instance \(E_T\), the common approval set of \(S\) is still \(A_S\), while its common disapproval set becomes \(D_S\setminus R\).
Let \(\alpha=|S|/n\), and define
\[
    \ell'
    :=
    \left\lfloor
    \alpha\cdot \bigl(|D_S\setminus R|+\min(k,|A_S|)\bigr)
    \right\rfloor .
\]
Thus \(S\) is \(\ell'\)-cohesive in the modified instance \(E_T\). Since \(S\) is \(\ell\)-cohesive in \(E\), we have
\[
\begin{aligned}
    \ell'
    &=
    \left\lfloor
    \alpha\cdot \bigl(|D_S|+\min(k,|A_S|)-|R|\bigr)
    \right\rfloor  \\
    &\geq
    \left\lfloor
    \alpha\cdot \bigl(|D_S|+\min(k,|A_S|)\bigr)
    \right\rfloor
    -
    \left\lceil \alpha|R| \right\rceil  \\
    &\geq
    \ell-\left\lceil \alpha|R|\right\rceil
    \geq
    \ell-|R|.
\end{aligned}
\]
Hence deleting the vetoes in \(R\) can reduce the formal claim of \(S\) by at most \(|R|\).

By the guarantee of \(\mathcal R\) applied to \(E_T\), group \(S\) obtains average satisfaction at least \(g(\ell')\) in the modified election. Since \(\mathcal R\) satisfies CVD at \(T\), no candidate in \(C\setminus T\) is selected in \(W\); in particular, no candidate in \(R\) is selected. Hence, when \(W\) is evaluated in the original election \(E\), the candidates in \(R\) add exactly \(|R|\) units of satisfaction to every voter in \(S\). Thus the average satisfaction of \(S\) in \(E\) is at least \(g(\ell')+|R|\). If \(\ell'\geq \ell\), this is at least \(g(\ell)\) by monotonicity of \(g\). Otherwise, using the assumed property of \(g\), \[ g(\ell')+|R| \geq g(\ell)-(\ell-\ell')+|R| \geq g(\ell), \] where the last inequality follows from \(\ell-\ell'\leq |R|\). Therefore, \(W\) gives \(S\) average satisfaction at least \(g(\ell)\) when evaluated in \(E\). \end{proof}

The following example shows that both rules we study in the symmetric setting violate counterfactual stability, in general.
\phragmenfailscs*
\begin{proof}
Let $C=\{a,b,c\}$, $N=\{1,2\}$ and $k=1$. The approval and disapproval sets are as follows: $A_1=\{a\}, D_1=\{b\}, A_2=\{b,c\}, D_2=\{a\}$.
We also fix the following tie-breaking:
(i) under Phragm\'en's rule, we execute veto actions before selection actions at equal times, and
(ii) under both rules, we break remaining ties lexicographically.
For any $T\subseteq C$, let $E_T$ be obtained by restricting vetoes to $T$.
As we will next show, the resulting winner for $E_T,$ under both rules is
\[
R(E_T)=
\begin{cases}
\{a\}, & a\notin T \text{ and } b\notin T,\\
\{b\}, & a\in T \text{ and } b\notin T,\\
\{c\}, & b\in T.
\end{cases}
\]
For PAV, this holds due to the PAV scores of the possible committees that are shown in the columns of \Cref{tab:pavcounter} for each possible choice of $T$.
For Phragmén’s rule, if $a\notin T$ and $b\notin T$, the outcome is ${a}$: both voters can afford their approved candidates, and ties are broken in favor of $a$.
If $a\in T$ and $b\notin T$, ties are resolved first in favor of a veto action, so $\neg a$ is purchased first; subsequently, the second voter purchases an approved candidate, namely $b$, according to the tie-breaking order.
If $a\in T$ and $b\in T$, the negations of the candidates are purchased first, leaving only candidate $c$ to be selected positively.
Finally, if $a\notin T$ and $b\in T$, the first voter purchases $\neg b$, after which the second voter can purchase a feasible approved candidate, which is $c$.
\begin{table}[t]
\centering
\begin{tabular}{c|cccc}
\toprule
Case for $T$ & $\emptyset$ & $\{a\}$ & $\{b\}$ & $\{c\}$ \\
\midrule
$a \notin T$, $b \notin T$ & 0 & 1   & 1   & 1   \\
$a \in T$, $b \notin T$    & 1 & 1   & 1.5 & 1.5 \\
$a \in T$, $b \in T$       & 2 & 1.5 & 1.5 & 2.5 \\
$a \notin T$, $b \in T$    & 1 & 1.5 & 1   & 2   \\
\bottomrule
\end{tabular}
\caption{PAV scores of all feasible positive committees under the
different possibilities for $T$.}
\label{tab:pavcounter}
\end{table}
We then consider the following cases:
\begin{itemize}
\item If $a\notin T$ and $b\notin T$, then $a\in R(E_T)$ and $a\notin T$;
      hence CVD is violated at $a$.
\item If $a\in T$ and $b\notin T$, then $b\in R(E_T)$ and $b\notin T$;
      hence CVD is violated at $b$.
\item If $b\in T$ and $a\notin T$, then $R(E_T)=\{c\}$, so
      $b\in T\setminus R(E_T)$, but
      $b\notin R(E_{T\setminus\{b\}})=\{a\}$; hence MVD is violated at $b$.
\item If $\{a,b\}\subseteq T$, then $R(E_T)=\{c\}$, so
      $a\in T\setminus R(E_T)$, but
      $a\notin R(E_{T\setminus\{a\}})=\{c\}$; hence MVD is violated at $a$.
\end{itemize}
 Hence, there exists no $T$ such that $R(E_T)$ satisfies both CVD and MVD at $T$.
\end{proof}
\paragraph{Game-theoretic interpretation of counterfactual stability.}
\label{app:veto-game}

there exists a set $T$ at which a rule (e.g.\ Phragm\'en or PAV) is stable.

\subsubsection{Counterfactual Tranformation for any Rule}

While we described in the main body a way to select $T$ for Phragm\'en, the same idea extends to any
rule $R$ by imposing a sequential veto-selection procedure on top of it. 

\medskip
\begin{mdframed}[linecolor=black, linewidth=0.3mm, innerleftmargin=10pt, leftmargin=10pt, rightmargin=10pt]
\textbf{Counterfactual Veto Selection for a rule $R$.}
Fix a resolute rule $R$ and an election $E=(N,C,(A_i)_{i\in N},(D_i)_{i\in N},k)$.
Fix an ordering $c_1,\dots,c_m$ of the candidates. Maintain an \emph{active veto set} $T\subseteq C$,
initially $T\gets C$, and process candidates in the order $c_1,\dots,c_m$.

At step $j$, having already fixed the decisions for $c_1,\dots,c_{j-1}$, decide whether to keep
vetoes against $c_j$ by a recursive counterfactual check: simulate the continuation of the
\emph{same veto-selection procedure} from the current state with vetoes against $c_j$ deleted
(equivalently, with active set $T\setminus\{c_j\}$). If $c_j$ is \emph{selected} in this recursive
continuation, then keep its vetoes (leave $c_j\in T$); otherwise delete them (set
$T\gets T\setminus\{c_j\}$).

Let $T^\star$ be the final active veto set, and output $R(E_{T^\star})$.
\end{mdframed}

\paragraph{Continuous removal of Disapprovals} {Instead of removing a candidate $c$ entirely from the set of counted disapprovals $T$, this adjustment can be made in a more gradual and controlled way. Indeed, the rules we consider naturally extend to settings with weighted votes. For example, in the case of PAV, this requires using the definition of harmonic numbers extended to fractional values, and interpreting the score that a voter assigns to a candidate $c$ as the harmonic value of the total weight assigned to the selected candidates, defined as in the work by \citet{lu2024approval}.
In this framework, $T$ can be defined as a function that assigns a weight to each candidate, and the disapproval associated with candidate $c$ is counted proportionally to the weight $T(c)$. The objective is then to determine the lowest weights that still ensure counterfactual stability.
While we initially described the process in terms of removing candidates $c$ either completely or not at all from $T$, this was mainly for the sake of exposition. In practice, we recommend gradually decreasing the weights of disapprovals rather than applying an all-or-nothing removal.}

The strategy for constructing a counterfactually stable set $T$ can also be extended to the removal of individual disapprovals and approvals\footnote{Intuitively, voters should not lose voting power when one of their approved candidates is selected, provided that this candidate would still be selected even if their approvals were removed.} and, even continuous removal of approvals and disapprovals. 

\subsection{Details on the Connections to General Feasibility Constraints}
\label{sec:app_base}
Our problem can be viewed as a special case of the framework on \emph{elections with general feasibility constraints}, a setting in which the concept of proportionality has been recently explored \cite{masavrik2024generalised}. 
In this section, we first establish the connection to this framework and the relevant known results.
We also revisit the proportionality notions introduced by \citet{masavrik2024generalised} to highlight their connection to our main concept of proportional representation (\Cref{def:axiom}), demonstrating the relevance and soundness of the axioms we examine. In the process, we also offer a reformulation of the central concept from \cite{masavrik2024generalised} to address a known limitation.

\citet{masavrik2024generalised} have already adapted PAV and Phragm\'en to that framework and proved that the two rules satisfy certain proportionality guarantees. Yet, they show that these guarantees apply if and only if the constraints form matroids.
Therefore, these results do not hold to our setting, as our so-constructed feasibility constraints do not have a matroid structure.\footnote{Consider, for instance, $k=2$ with the outcomes $\{a,b,\neg c\}$ and $\{c,a\}$. Neither $\{c,a,b\}$ nor $\{c,a,\neg a\}$ forms a feasible set.} Nevertheless, our construction implicitly introduces a key ballot restriction: no voter can approve both a candidate $c$ and $\neg c$ simultaneously. This structural property of voters' preferences allowed us to circumvent the impossibility results of~\citet{masavrik2024generalised} and establish the results in \Cref{sec:pav,sec:phragmen}.

We now recall the axiom of Base Extended Justified Representation (Base EJR) \citet{masavrik2024generalised}. For that, we highlight that it gives the guarantees that the prominent axiom of EJR~\citep{aziz2017justified} would give, when we restrict attention to instances of our framework with $D_i=\emptyset, \forall i\in V.$ 
We also note that this axiom is always satisfiable and particularly strong: apart from EJR for committee elections, 
it also implies proportionality for cohesive groups in the model of public decisions~\citep{sko-gor:proportional_public_decisions} and strong EJR in the context of sequential decision making~\citep{cha-goe-pet:seq-decision-making}.
It roughly says that a set of voters $S$ deserves $\ell$ candidates if they can complete any reasonable selection made by the remaining voters, $T,$ with $\ell$ commonly approved candidates. Let $\calF$ be the collection of feasible outcomes. In our case:
\begin{align}
\label{eq:constraints}
\calF = \{W \subseteq C \cup \neg C \text{~such that~} |W \cap C| \leq k \text{~and~} \neg(W\cap C) \cap (W \cap \neg C) = \emptyset\} \text{.}
\end{align}
 \vspace{-0.6cm}
\begin{definition}[\textbf{Base Extended Justified Representation (Base EJR)}]
    Consider an election $E=(C,V,k,B).$ Given a positive integer $\ell,$ we say that a set of voters $S\subseteq V$ is $\ell$-cohesive if for each feasible solution $T \in \calF$ either there exists a set $X\subseteq A_S$ with $|X|\geq \ell$ such that $T\cup X$ remains feasible, or $\nicefrac{|S|}{n}>\nicefrac{\ell}{|T|+\ell}.$ An outcome $W$ is said to provide Base EJR for $E$ if for every $\ell$-cohesive set of voters $S$ there exists a voter $i$ with a satisfaction $u_i(W)$ of at least $\ell$:
        \begin{align*}
    u_i(W) = |A_i \cap W| + |\neg D_i \cap  W| \geq \ell \text{.}
    \end{align*}
        A rule $\mathcal{R}$ satisfies Base EJR if for every election $E$ its winning outcome provides Base EJR. \hfill $\lrcorner$
\end{definition}

The definitions of Base EJR, as well as the analogous definition of Base PJR \cite{masavrik2024generalised}, require considering all possible sets of $T$, which makes it difficult to interpret in terms of the actual number of representatives a group $S$ is entitled to; denoted by $\vclaim(S).$
At what follows we offer a lemma that addresses this limitation by providing a closed formula for this value. Specifically, given an election and a voter set $S,$ our result allows for a direct computation of the satisfaction that voters in $S$ are guaranteed under the base axioms.
The formula is complex because it precisely accounts for all edge cases, particularly those where the sizes of sets $A_S$ and $D_S$ are disproportionate.

\begin{restatable}{proposition}{propBaseEjrAndEjr}\label{prop:base-ejr-and-ejr}
Consider an election $E$. A set of voters $S\subseteq V$ is $\ell$-cohesive if:
\begin{equation*}\label{eq:base-ejr-formula}
    \ell \leq \begin{cases}
        |D_S| - k, &\mathrm{if~~} \frac{n}{n-|S|}k \leq |D_S|.\\[7pt]
        \frac{|S|}{2n-|S|}(|D_S| + k), &\mathrm{if~~} \frac{n-|S|}{n}k \leq |D_S| \leq \frac{n}{n-|S|} k \mathrm{~~and~~} \frac{2n-|S|}{n} |A_S|  + \frac{n-|S|}{n} |D_S| \geq k. \\[7pt]
        \frac{|S|}{n}k, &\mathrm{if~~} |A_S|+ |D_S| \geq k \mathrm{~~and~~} |D_S| \leq \frac{n-|S|}{n}k \mathrm{~~and~~} |A_S| \leq m - \frac{n-|S|}{n}k. \\[7pt]
        |A_S| + k - m, &\mathrm{if~~} |A_S| + |D_S| \geq k \mathrm{~~and~~} |D_S| \leq \frac{n-|S|}{n}k \mathrm{~~and~~}\\[1pt]
        &\phantom{\mathrm{if~~}}  m - \frac{n-|S|}{n}k \leq |A_S| \mathrm{~~and~~} |A_S| + k - m \leq \frac{|S|}{n}(|A_S|+|D_S|). \\[4pt]
        \frac{|S|}{n}(|D_S| + |A_S|), &\mathrm{otherwise.}
    \end{cases}
\end{equation*}
All the bounds are tight up to an additive term of $\nicefrac{|S|}{n-|S|}$.
\end{restatable}

\begin{proof}
    \Crefname{equation}{formula}{formulae}
    \Crefname{equation}{Formula}{Formulae}

    Consider an election $E$ and a group $S\subseteq N$. Let us denote by $\vclaim(S)$ the upper bound of $\ell$ for $S$ in the statement of the lemma. To prove the statement, we will show that each group of voters $S\subseteq V$ is $\vclaim(S)$-cohesive and is not $(\vclaim(S)+1)$-cohesive. 
    
    Consider a feasible subset of candidates $T\subseteq C$ and let $T^+$ and $T^-$ denote the parts of $T$ that consist of the positive and negative candidates, respectively. 
    If we have that $\frac{|S|}{n} > \frac{\vclaim(S)}{\vclaim(S) + |T|}$, then $S$ is entitled to satisfaction $\vclaim(S)$. From now on let us assume that the opposite inequality holds, which is equivalent to: \begin{equation}\label{eq:bound-on-Tapp}
        |T| \leq \frac{n-|S|}{|S|}\cdot \vclaim(S).
    \end{equation}

    Moreover, let us assume that the expression $\frac{n-|S|}{|S|}\cdot \vclaim(S)$ is integral, that is, it is possible to find set $T$ of exactly this size. Whenever it is true, the estimations for $\vclaim(S)$ will be tight. Otherwise, since the space of possible sets $T$ is smaller, the actual value of $\vclaim$ might be larger by less than:
    \begin{align}\label{eq:estimation-upper-bound2}
    \begin{split}
        \vclaim(S) &- \frac{|S|}{n-|S|}\cdot \left\lfloor \frac{n-|S|}{|S|} \cdot \vclaim(S) \right\rfloor \\ &= \frac{|S|}{n-|S|}\cdot \left(    \frac{n-|S|}{|S|} \cdot \vclaim(S) - \left\lfloor \frac{n-|S|}{|S|} \cdot \vclaim(S) \right\rfloor \right) < \frac{|S|}{n-|S|}.
    \end{split}
    \end{align}

    So we want to show that for any choice of such a set $T$, there is a set $X\subseteq A_S\cup \neg D_S$ that can be proposed by group $S$ such that $T\cup X$ is feasible and $|X|= \vclaim(S)$ and there is no such larger set $X$. This will give us the desired result, because such a set $T$ certifies that group $S$ is not $(\vclaim(S)+1)$-cohesive as \Cref{eq:bound-on-Tapp} is also satisfied for any $\ell>\vclaim(S)$.
    
    Now, subject to the constraint of \Cref{eq:bound-on-Tapp} we will 
    choose $T$ adversarially, so as to minimize the size of the largest such set $X$.

    Given a set $T$, group $S$ should be able to propose a set of candidates $X\subseteq A_S\cup \neg D_S$ of size $\vclaim(S)$ (but not any larger) such that $T\cup X$ is feasible. 
    Due to the committee size constraint, the set $X$ contains at most $k-|T^+|$ candidates from $A_S.$ Furthermore, $X\cup T$ must not include both $a$ and $\neg a$ for any $a\in C$. Hence the maximum sized set $X$ satisfies $|X\cap \neg D_S|\leq | D_S\setminus T^+|$ and $|X\cap A_S|\leq |A_S \setminus \neg T^-|$.
    From these inequalities, we obtain the following formula for the maximum size of set $X$ that $S$ can propose so that $T\cup X$ is feasible:
    \begin{align*}
    |D_S\setminus T^+| + &\min(|A_S \setminus \neg T^-|, k - |T^+|)\\
    =    |D_S| - |T^+| + |T^+ \setminus D_S| + &\min(|A_S|, k - |T^+| + |A_S \cap \neg T^-|) - |A_S\cap \neg T^-|.\notag
    \end{align*}
    
In addition to the size constraint (\Cref{eq:bound-on-Tapp}), we now impose further restrictions on $T$ and show that these do not decrease the size of the largest set $X\subseteq A_S\cup \neg D_S$ for which $X\cup T$ is feasible. We can assume that $T$ has the following properties without loss of generality:\; $T^+\cap A_S =\emptyset$ and $T^-\subseteq \neg A_S.$ To justify this claim, observe that
    removing a candidate from $T$ does not affect the satisfiability of \Cref{eq:bound-on-Tapp}. 
    Removing the candidates from $T^+\cap A_S$ does not increase the maximum possible size of $X$ (i.e., the number of ``slots'' in $X$ available for the candidates from $A_S \cup \neg D_S$). The same holds for removing the candidates from $T^-\setminus \neg A_S.$
    Moreover, if for $T$ there is a set $X\subseteq A_S \cup D_S$ of size at least $\vclaim(S)$, then this also holds for any subset of $T$.

    Therefore, we have that $T^-$ contains only candidates from $\neg A_S,$ and, $|A_S\cap \neg T^-| = |\neg T^-| = |T^-|$. Hence, the maximum size of $X$ is:
    \begin{align}\label{eq:main-base-ejr}
    |D_S| - |T^+| + |T^+ \setminus D_S| + &\min(|A_S|, k - |T^+| + |A_S \cap \neg T^-|) - |A_S\cap \neg T^-|\notag\\
        = |D_S| - |T^+| + |T^+ \setminus D_S| + &\min(|A_S|, k - |T^+| + |T^-|) - |T^-| \notag \\ = |D_S| + |T^+ \setminus D_S| + &\min(|A_S|, k - |T^+| + |T^-|) - |T|.
    \end{align}
    
    Let us now check for which sets $T$ the \Cref{eq:main-base-ejr} is minimal.
    Any candidate, positive or negative, included in $T$, contributes $-1$ to the last term $- |T|$.
    Among those, only positive candidates included in $T$ contribute $-1$ to the term $k - |T^+| + |T^-|$ and thereby potentially reduce the size of $X$ (this is in contrast to negative candidates, which increase this term).
    Finally, only positive candidates in $D_S$ additionally contribute $0$ to $|T^+\setminus D_S|$ if included in $T^+$ while all other positive candidates contribute $1$.
    Thus, to minimize the formula, we always prefer to add candidates from $D_S$ to $T^+$ instead of the ones from $C\setminus D_S$ (subject to \Cref{eq:bound-on-Tapp} and committee constraint $k$).
    So, if possible, we choose $T$ to be of size $\frac{n-|S|}{|S|}\cdot \vclaim(S)$ and $T=T^+=D_S.$ If $|D_S|$ or $k$ are too small (that is, smaller than $\frac{n-|S|}{|S|}\cdot \vclaim(S)$, so that using up to the $k$ maximum possible candidates from $D_S$ is insufficient), we will sometimes fill the remaining slots in $T$ with candidates from $C\setminus D_S\setminus A_S$ to $T^+$ or the ones from $\neg A_S$ to $T^-.$ 
    The remainder of the proof requires an extensive case analysis. 

\begin{description}
    \item[Case 1: \textnormal{Assume that $|D_S| \geq \frac{n}{n-|S|}\cdot k$.}] ~\\
    In this case we have that $\vclaim(S) = |D_S| - k$. The condition in the assumption can be rewritten as:
    \begin{align*}
        |D_S| \geq \frac{n}{n-|S|}\cdot k \iff |D_S| - k \geq \frac{|S|}{n-|S|}\cdot k \iff \frac{n-|S|}{|S|}\cdot \vclaim(S) \geq k.
    \end{align*}
    Then it is possible to take $T = T^+ \subseteq D_S$ such that $T^+$ has the maximal possible size of $k$. As argued in the main text, this minimizes \Cref{eq:main-base-ejr}.
        Then we can transform \Cref{eq:main-base-ejr} as:
\begin{align*}
    |D_S| + \min(|A_S|, 0) - k =|D_S| - k = \vclaim(S),
\end{align*}
which shows that indeed Base EJR gives precisely a guarantee of $\vclaim(S)$ in this case. ~\\

\item[Case 2: \textnormal{Assume that $\frac{n-|S|}{n}\cdot k \leq |D_S| \leq \frac{n}{n-|S|}\cdot k$ and $\frac{2n-|S|}{n} |A_S|  + \frac{n-|S|}{n} |D_S| \geq k$.}] ~\\
In this case we have that $\vclaim(S) = \frac{|S|}{2n-|S|}\cdot (|D_S|+k)$. 
The assumption $\frac{n-|S|}{n}\cdot k \leq |D_S| \leq \frac{n}{n-|S|}\cdot k$ implies both $\frac{n-|S|}{|S|}\cdot \vclaim(S)\leq \frac{n-|S|}{2n-|S|}(\frac{n}{n-|S|}k+\frac{n-|S|}{n-|S|}k)$ and $\frac{n-|S|}{|S|}\cdot \vclaim(S)\leq \frac{n-|S|}{2n-|S|}(\frac{n-|S|}{n-|S|}D_S+\frac{n}{n-|S|}D_S)$. Therefore,
\[\frac{n-|S|}{|S|}\cdot \vclaim(S) \leq |D_S|, \quad \text{ and } \quad \frac{n-|S|}{|S|}\cdot \vclaim(S) \leq k.\]
Hence, it is possible to take $T = T^+ \subseteq D_S$ such that $T$ has the maximal possible size of $\frac{n-|S|}{|S|}\cdot \vclaim(S)$. Now we can rewrite \Cref{eq:main-base-ejr} as:
\begin{align}\label{eq:case2-max-X}
    |D_S| + \min(|A_S|, k - \frac{n-|S|}{|S|}\cdot \vclaim(S)) - \frac{n-|S|}{|S|}\cdot \vclaim(S).
\end{align}
Then, from the assumptions of the considered case we obtain
$$\frac{2n-|S|}{n} |A_S|  + \frac{n-|S|}{n} |D_S| \geq k,$$
which can be equivalently written as:
\begin{align*}
     &|A_S| \geq \frac{k - \frac{n-|S|}{n}|D_S|}{\frac{2n-|S|}{n}} \\&\iff |A_S| \geq k - \frac{n-|S|}{2n-|S|}(|D_S| + k) \\ &\iff |A_S| \geq k - \frac{n-|S|}{|S|}\cdot \frac{|S|}{2n-|S|}(|D_S| + k) \\ &\iff |A_S| \geq k - \frac{n-|S|}{|S|}\vclaim(S).
\end{align*}
So, $\min(|A_S|,k - \frac{n-|S|}{|S|}\vclaim(S))=k - \frac{n-|S|}{|S|}\vclaim(S)$ in \Cref{eq:case2-max-X}, which can be further simplified as follows
\begin{align*}
    &|D_S| + k - \frac{n-|S|}{|S|}\cdot \vclaim(S) - \frac{n-|S|}{|S|}\cdot \vclaim(S) \\= &|D+S|+k-\frac{2(n-|S|)}{|S|}\frac{|S|}{2n-|S|}(|D_S|+k)=\frac{|S|}{2n-|S|}\cdot (|D_S|+k) = \vclaim(S),
\end{align*}
which shows that indeed Base EJR gives a guarantee of $\vclaim(S)$ in this case. ~\\

\item[Case 3: \textnormal{Assume that $|A_S| + |D_S| \geq k$ and $|D_S| \leq \frac{n-|S|}{n}\cdot k$ and $|A_S| \leq m - \frac{n-|S|}{n}\cdot k$.}] ~\\
In this case, we have that $\vclaim(S) = \frac{|S|}{n} k$. Hence, $|D_S|\leq \frac{n-|S|}{n}\cdot k = \frac{n-|S|}{|S|}\frac{|S|}{n}\cdot k\leq \frac{n-|S|}{|S|}\cdot \vclaim(S)$. For the purpose of analysis consider $T=T^+ = D_S,$ we will enlarge $T$ thereafter. Under this assumption the term $\min(|A_S|, k-|T^+|+|T^-|)$ in \Cref{eq:main-base-ejr} equals $k-|D_S|$ using that $|A_S| + |D_S| \geq k$. 
We now add further $x = \frac{n-|S|}{n}\cdot k - |D_S|$ to $T$ so as to minimize \Cref{eq:main-base-ejr}.
Since for our temporary choice of $T$ it holds that $k-|T^+|+|T^-|\leq |A_S|,$ it is true that the $min$ term can be further reduced by adding $x$ positive candidates from $C \setminus D_S \setminus A_S$ to $T^+$, upon which 
term  $\min(|A_S|, k-|T^+|+|T^-|)$ decreases by $x$ while $|T^+\setminus D_S|$ increases by $x$ and the part $-|T|$  decreases by $x$.
 Overall, \Cref{eq:main-base-ejr} decreases by $x$. Note that if we increased the size of $T$ in any other way (i.e., by adding candidates to $T^-$) the decrease would be smaller, so that our choice of $T$ minimizes \Cref{eq:main-base-ejr}.
It remains to justify that there are at least $x=\frac{n-|S|}{n}\cdot k - |D_S|$ candidates in $C\setminus D_S\setminus A_S$ or equivalently, whether $m - |D_S| - |A_S| \geq \frac{n-|S|}{n}\cdot k - |D_S|$. The latter condition is equivalent to $|A_S| \leq m - \frac{n-|S|}{n}\cdot k$, which holds by assumption.
Finally, we simplify \Cref{eq:main-base-ejr} as:
\begin{align*}
&|D_S|+|T^+\setminus D_S|+\min(|A_S|, k-|T^+|+|T^-|)-|T|=\\&|D_S|+\frac{n-|S|}{|S|}\vclaim(S) - |D_S|+k-|D_S|-\left(\frac{n-|S|}{|S|}\vclaim(S) - |D_S|\right)- \frac{n-|S|}{|S|}\vclaim(S)
    =\\&\frac{n-|S|}{|S|}\vclaim(S) +  k - \frac{n-|S|}{|S|}\vclaim(S) - \frac{n-|S|}{|S|}\vclaim(S)=k - \frac{n-|S|}{|S|}\vclaim(S) =\vclaim(S),
\end{align*}
which shows that indeed Base EJR gives a guarantee of $\vclaim(S)$ in this case.  ~\\

\item[Case 4: \textnormal{Assume that $|A_S| + |D_S| \geq k$ and $|D_S| \leq \frac{n-|S|}{n}\cdot k$ and $m - \frac{n-|S|}{n}\cdot k \leq |A_S|$.}] ~\\
In this case we can have that either $\vclaim(S)=|A_S|+k-m$ or $\vclaim(S) = \frac{|S|}{n}(|A_S|+|D_S|)$, depending on which value is smaller. Alternatively, we can say that $\vclaim(S) = \min(|A_S|+k-m, \frac{|S|}{n}(|A_S|+|D_S|))$.
We have that:
\begin{align*}
    \frac{n-|S|}{|S|}(|A_S| + k - m) &\geq \frac{n-|S|}{|S|}(m - \frac{n-|S|}{n}\cdot k + k - m) = \frac{n-|S|}{n}\cdot k \geq |D_S|,
\end{align*}
\begin{align*}
    \frac{n-|S|}{n}(|A_S| + |D_S|) &\geq \frac{n-|S|}{n}k \geq |D_S|,
\end{align*}
which means that $|D_S| \leq \frac{n-|S|}{|S|}\vclaim(S)$ and, as in Case 3, we first include all candidates from $D_S$ in $T$.
Now, the question is whether set $T$ should be completed by the candidates from $C\setminus A_S \setminus D_S$ or by the candidates from $\neg A_S.$ Note that from our assumptions, we have that:
\begin{align*}
    m - |A_S| &\leq \frac{n-|S|}{n}k \leq k,\\
    m - |A_S| &\leq \frac{n-|S|}{n}k \leq \frac{n-|S|}{n}(|A_S|+|D_S|),\\
    m - |A_S| &\leq \frac{n-|S|}{n}k \iff \frac{n}{|S|}m - \frac{n}{|S|}|A_S| \leq \frac{n-|S|}{|S|}k \iff  m - |A_S| \leq \frac{n-|S|}{|S|}(|A_S| + k - m),
\end{align*}
which means that $m-|A_S| \leq \min(k, \frac{n-|S|}{|S|\vclaim(S)})$ and it is possible to include all the candidates from $C \setminus A_S \setminus D_S$ in $T$. On the other hand, we have that $|A_S| + |D_S| \geq \frac{|S|}{n}(|A_S| + |D_S|)$ and $|A_S| + |D_S| \geq k \geq |A_S| + k - m$, which means that it is not possible to include all the candidates from $\neg A_S$ in $T$.

Looking at the \Cref{eq:main-base-ejr}, we can prove the following claim:

\begin{claim}
\label{claim7}
    An optimal set $T$ minimizing \Cref{eq:main-base-ejr}, either contains all the candidates from $C\setminus A_S\setminus D_S,$ or none of them.
\end{claim}

\begin{proof}[Proof of \Cref{claim7}]\, 
    Indeed, consider a set $T$ in which there is some $x$ ($0 < x < m - |A_S| - |D_S|$) candidates from $C\setminus A_S\setminus D_S$ and some $y$ candidates from $\neg A_S.$ We will show that $T$ does not minimize \Cref{eq:main-base-ejr}.

    We know that $y < |A_S|$ (as we noted above, it is not possible to include all the candidates from $\neg A_S$ in $T$). First of all, consider the case when $y=0$. Then, as noted above, $|T| < m-|A_S| \leq \frac{n-|S|}{|S|}\vclaim(S)$ and it is possible to add a candidate from $\neg A_S$ to $T$. From the perspective of minimizing \Cref{eq:main-base-ejr}, such an operation clearly is (weakly) profitable---it always decreases the value of the part $-|T|$ by one, and it can increase the value of part $\min(|A_S|, k - |T^+| + |T^-|)$ only by at most one. Further, let us assume that $y > 0$.

Observe now that the part $\min(|A_S|, k - |T^+| + |T^-|)$ is either equal to $|A_S|$ or to $k - |T^+| + |T^-|$. In the first case, we consider $T'$ obtained from $T$ by making a swap between $c\in T \cap (C\setminus A_S \setminus D_S)$ and a candidate $c'\in A_S \setminus T$. In the second case, we consider $T'$ obtained from $T$ by making a swap between a candidate $c\in (C \setminus A_S \setminus D_S) \setminus T$ and a candidate $c'\in T \cap A_S.$ It is straightforward to check that in both cases the value of \Cref{eq:main-base-ejr} under $T'$ decreases, which proves that $T$ does not minimize \Cref{eq:main-base-ejr}.
\end{proof}

Hence, the optimal $T$ in this case either contains all the candidates from $D_S$ and $\frac{n-|S|}{|S|}\vclaim(S) - |D_S|$ candidates from $\neg A_S$ (let us denote such $T$ by $T_1$) or all the candidates from $C\setminus A_S$ and $\frac{n-|S|}{|S|}\vclaim(S) - (m-|A_S|)$ candidates from $\neg A_S$ (let us denote such $T$ by $T_2$. In the first case, \Cref{eq:main-base-ejr} can be transformed as follows:
\begin{align}\label{eq:base-ejr-t1}
    |D_S| + \min(|A_S|, k - |D_S| + \frac{n-|S|}{|S|}\vclaim(S) - |D_S|) - \frac{n-|S|}{|S|}\vclaim(S).
\end{align}

In the second case, \Cref{eq:main-base-ejr} can be transformed as follows:
\begin{align}\label{eq:base-ejr-t2}
    m - |A_S| + \min(|A_S|, k - (m - |A_S|) + \frac{n-|S|}{|S|}\vclaim(S) - (m - |A_S|)) - \frac{n-|S|}{|S|}\vclaim(S)\notag \\
    = m - |A_S| + \min(|A_S|, k - 2m + 2|A_S| + \frac{n-|S|}{|S|}\vclaim(S)) - \frac{n-|S|}{|S|}\vclaim(S).
\end{align}

Let us now consider the following subcases:~\\

\item[Subcase 4.1:]\,Assume that $|A_S| + k - m \geq \frac{|S|}{n}(|A_S|+|D_S|)$.~\\
In this subcase, we have that $\vclaim(S) = \frac{|S|}{n}(|A_S|+|D_S|)$. From our assumption we have that:
\begin{align*}
\begin{split}
    & \frac{|S|}{n}(|A_S|+|D_S|) \leq |A_S| + k - m \\
    \iff &\frac{|S|}{n}(|A_S|+|D_S|) + |D_S| \leq |A_S| + |D_S| + k - m \\
    \implies &\frac{|S|}{n}|A_S| + \frac{n+|S|}{n}|D_S| \leq k\\
    \iff &|A_S| + 2|D_S| \leq k + \frac{n-|S|}{n}(|A_S|+|D_S|)\\
    \iff &|A_S| \leq k - |D_S| + \frac{n-|S|}{|S|}\vclaim(S) - |D_S|. 
\end{split}
\end{align*}
Hence, taking $T=T_1$, we can transform \Cref{eq:base-ejr-t1} as:
\begin{align*}
    |D_S| + |A_S| - \frac{n-|S|}{n}(|A_S|+|D_S|) = \frac{|S|}{n}(|A_S|+|D_S|) = \vclaim(S),
\end{align*}
and when we take $T=T_2$, then, depending on the value of the $\min$ part, we can transform \Cref{eq:base-ejr-t2} either as:
\begin{align*}
    m - |A_S| + |A_S| - \frac{n-|S|}{n}(|A_S| + |D_S|) \geq \frac{|S|}{n}m \geq \vclaim(S)
\end{align*}
or as:
\begin{align*}
    m - |A_S| + k - 2m + 2|A_S| + \frac{n-|S|}{|S|}\vclaim(S) - \frac{n-|S|}{|S|}\vclaim(S) \\ = |A_S| + k - m \geq \frac{|S|}{n} (|A_S| + |D_S|) = \vclaim(S).
\end{align*}
In all cases group $S$ has a guarantee of at least $\vclaim(S)$ and for $T=T_1$, the inequality is tight, which shows that indeed Base EJR gives a guarantee of $\vclaim(S)$ in this subcase.~\\

\item[Subcase 4.2:]\,Assume that $|A_S| + k - m \leq \frac{|S|}{n}(|A_S|+|D_S|)$.~\\
In this subcase we have that $\vclaim(S) = |A_S|+k-m$. Note that our assumption, together with the fact that $|A_S|+|D_S| \leq m$, implies that $|A_S|+k-m \leq \frac{|S|}{n}m$. From that we further obtain:
\begin{align*}
    &\frac{|S|}{n}m \geq |A_S|+k-m \\
    \iff &\frac{n+|S|}{n}m -k \geq |A_S|\\
    \iff & |A_S| \geq k + 2|A_S| -  \frac{n+|S|}{n}m\\
    \iff &|A_S| \geq  k-2m+2|A_S|+\frac{n-|S|}{n}m\\
    \implies & |A_S| \geq  k-2m+2|A_S|+\frac{n-|S|}{|S|}(|A_S|+k-m)\\
    \iff & |A_S| \geq  k-2m+2|A_S|+\frac{n-|S|}{|S|}\vclaim(S).
\end{align*}
Therefore, taking $T=T_2$, we can transform \Cref{eq:base-ejr-t2} as follows:
\begin{align*}
    m - |A_S| + k - 2m + 2|A_S| + \frac{n-|S|}{|S|}\vclaim(S) - \frac{n-|S|}{|S|}\vclaim(S) = |A_S| + k - m = \vclaim(S).
\end{align*}
If we take $T=T_1$, then, depending on the value of the $\min$ part, we can transform \Cref{eq:base-ejr-t1} either as:
\begin{align*}
    &|D_S| + |A_S| - \frac{n-|S|}{|S|}(|A_S|+k-m) \geq |D_S| + |A_S| - \frac{n-|S|}{n}(|A_S|+|D_S|) 
    \\&= \frac{|S|}{n}(|A_S|+|D_S|)\geq |A_S|+k-m = \vclaim(S),
\end{align*}
or as:
\begin{align*}
    |D_S| + k - 2|D_S| + \frac{n-|S|}{|S|}\vclaim(S) - \frac{n-|S|}{|S|}\vclaim(S) = k - |D_S| \geq |A_S| + k - m = \vclaim(S).
\end{align*}
In all cases group $S$ has a guarantee of at least $\vclaim(S)$ and for $T=T_2$, the inequality is tight, which shows that indeed Base EJR gives a guarantee of $\vclaim(S)$ in this subcase. ~\\

\item[Case 5: \textnormal{Assume that none of the conditions examined in the previous cases hold.}] ~\\
In this case we have that $\vclaim(S) = \frac{|S|}{n}\cdot (|A_S| + |D_S|)$. We start by proving that now that at least one of the following is true:
\begin{equation}\label{eq:base-ejr-remaining-cases}
\begin{split}
    &\left(\frac{2n-|S|}{n}|A_S| + \frac{n-|S|}{n}|D_S| \leq k \mathrm{~~and~~} |A_S| \leq \frac{|S|}{n-|S|} |D_S|\right) \mathrm{~~or~~} \\
    &\left(|A_S| + |D_S| \leq k \mathrm{~~and~~} |A_S| \geq \frac{|S|}{n-|S|} |D_S|\right).
\end{split}
\end{equation}

Suppose first that $\frac{2n-|S|}{n}|A_S| + \frac{n-|S|}{n}|D_S| > k$. Then, since the conditions for Cases 1 and 2 are not satisfied, it has to hold that $|D_S| < \frac{n-|S|}{n}k$. Since the conditions for Cases 3 and 4 are not satisfied, it has to hold that $|A_S|+|D_S| < k$. But then $|A_S| + |D_S| < \frac{2n-|S|}{n}|A_S| + \frac{n-|S|}{n}|D_S|$, from which we obtain $|A_S| > \frac{|S|}{n-|S|} |D_S|$. Eventually, the second part of \Cref{eq:base-ejr-remaining-cases} is satisfied.

Suppose now that $\frac{2n-|S|}{n}|A_S| + \frac{n-|S|}{n}|D_S| \leq k$ and $|A_S| > \frac{|S|}{n-|S|} |D_S|$. Then we have that $k \geq \frac{2n-|S|}{n}|A_S| + \frac{n-|S|}{n}|D_S| = |A_S| + \frac{n-|S|}{n}|A_S| + \frac{n-|S|}{n}|D_S| > |A_S| + \frac{|S|}{n}|D_S| + \frac{n-|S|}{n}|D_S|  = |A_S| + |D_S|$. Therefore, once again, we have that the second part of \Cref{eq:base-ejr-remaining-cases} is satisfied.

We will now split the further analysis into two subcases, based on \Cref{eq:base-ejr-remaining-cases}.
\item[Subcase 5.1: \textnormal{Assume that $\frac{2n-|S|}{n}|A_S| + \frac{n-|S|}{n}|D_S| \leq k$ and $|A_S| \leq \frac{|S|}{n-|S|} |D_S|$}.] ~\\

In this subcase, the following two relations hold:
\begin{align*}
    |A_S| + |D_S| \leq k \implies \frac{n-|S|}{n} (|A_S| + |D_S|) \leq k \iff \frac{n-|S|}{|S|} \vclaim(S) &\leq k\\
    |A_S| \leq \frac{|S|}{n-|S|} |D_S| \iff \frac{n-|S|}{n} (|D_S| + |A_S|) \leq |D_S| \iff \frac{n-|S|}{|S|} \vclaim(S) &\leq |D_S|
\end{align*}
Hence, in this case it is possible to take $T = T^+ \subseteq D_S$ such that $|T|=\frac{n-|S|}{|S|}\vclaim(S)$. The \Cref{eq:main-base-ejr} then can be written as:
\begin{equation*}
    |D_S| + \min(|A_S|, k - \frac{n-|S|}{|S|}\vclaim(S)) - \frac{n-|S|}{|S|}\vclaim(S).
\end{equation*}
Since $\frac{2n-|S|}{n}|A_S| + \frac{n-|S|}{n}|D_S| \leq k \iff |A_S| \leq k - \frac{n-|S|}{n}(|A_S| + |D_S|) \iff |A_S| \leq k - \frac{n-|S|}{|S|}\vclaim(S)$, we may further transform \Cref{eq:main-base-ejr} as:
\begin{equation*}
    |D_S| + |A_S| - \frac{n-|S|}{n}(|A_S|+|D_S|) = \frac{|S|}{n}(|A_S|+|D_S|) = \vclaim(S),
\end{equation*}
which shows that indeed Base EJR gives a guarantee of $\vclaim(S)$ in this case.
\item[Subcase 5.2: \textnormal{Assume that $|A_S| + |D_S| \leq k$ and $|A_S| \geq \frac{|S|}{n-|S|} |D_S|$}.] ~\\
In this subcase the following equivalence holds:
\begin{equation*}
    |A_S| \geq \frac{|S|}{n-|S|} |D_S| \iff |D_S| \leq \frac{n-|S|}{n}(|D_S|+|A_S|) \iff |D_S| \leq \frac{n-|S|}{|S|}\vclaim(S).
\end{equation*}
Hence, we can add all the candidates from $D_S$ to $T$. Then, \Cref{eq:main-base-ejr} becomes:
\begin{equation*}
    |D_S| + \min(|A_S|, k - |D_S|) - |D_S|
\end{equation*}

However, we can still add $\frac{n-|S|}{|S|}\vclaim(S) - |D_S|$ candidates more to $T$. Since we assumed that $|A_S|+|D_S|\leq k,$ it also holds that $\min(|A_S|, k-|D_S|) = |A_S|$ and we can add more candidates to $T^-$ without increasing this part of the expression. Note that $|A_S|+|D_S|=\frac{n}{|S|}\vclaim(S)\geq \frac{n-|S|}{|S|}\vclaim(S)$, hence, we can add $\frac{n-|S|}{|S|}\vclaim(S) - |D_S|$ candidates from $\neg A_S$ to $T$. After that, we may further transform \Cref{eq:main-base-ejr} as:
\begin{equation*}
    u_i(W) = |D_S| + |A_S| - \frac{n-|S|}{|S|}\vclaim(S) = \vclaim(S),
\end{equation*}
which shows that indeed Base EJR gives a guarantee of $\vclaim(S)$ in this case.
\end{description} 
To conclude the case analysis, we can see than in each case Base EJR provides the guarantee of exactly $\vclaim(S)$ for the worst-case adversarial set, which completes the proof.
\end{proof}

While \Cref{eq:base-ejr-formula} appears to be quite complex, its core logic can be captured by a much simpler equation. The estimation we provide below not only clarifies the guarantees provided by Base EJR in our context, but it also confirms that these are well aligned with our intuitive understanding of proportionality as already formulated in \Cref{def:axiom}.
Note that the additive term of $\nicefrac{|S|}{n-|S|}$ follows from the potential non-tightness of the bounds in \Cref{prop:base-ejr-and-ejr}, due to discretization via rounding.

\begin{restatable}{proposition}{lemBaseEjrEstimation}\label{lem:base-ejr-estimation}
	Consider an election $E$ and fix any $\ell$-cohesive group of voters $S$. For each $k$-element subset $W \subseteq C,$ there exists a voter $i\in S$ such that $u_i(W) \geq \ell$ or
\begin{align*}
\ell < \frac{|S|}{n}\left(|D_S| + \min(k, |A_S|)\right) + \frac{|S|}{n-|S|} \text{.}
\end{align*}
\end{restatable}

\begin{proof}
The proof strategy is as follows: we consider the cases in the expression of the statement of \Cref{prop:base-ejr-and-ejr} to show that all of them are less than or equal $\frac{|S|}{n}\left(|D_S| + \min(k, |A_S|)\right)$. After that, it is sufficient to use \Cref{eq:estimation-upper-bound2} to prove the statement of the lemma.

First, observe that the voters from $S$ will always have the satisfaction of at least $|D_S| - k$ no matter what is the result of the election. Similarly, if we select $k$ candidates then, we will not select $(m-k)$ of them. Thus, the voters from $S$ will have at least the satisfaction of $|A_S| + k - m$.
Hence, we can focus on the remaining three cases only, namely the second, third, and the last one.

Consider the second case. Clearly, it always holds that:
\begin{align*}
\frac{2n-|S|}{n} k  + \frac{n-|S|}{n} |D_S|=k + \frac{n-|S|}{n}( k  +  |D_S|) \geq k.
\end{align*}
Consequently, since, by assumption, it also holds that $\frac{2n-|S|}{n} |A_S|  + \frac{n-|S|}{n} |D_S| \geq k,$ 
we get that:
\begin{align*}
\frac{2n-|S|}{n} \min(k, |A_S|)  + \frac{n-|S|}{n} |D_S| \geq k.
\end{align*}
After reformulation, we get that:
\begin{align*}
\frac{2n-|S|}{n} \min(k, |A_S|)  + \frac{2n-|S|}{n} |D_S| \geq |D_S| + k\text{,}
\end{align*}
from which our statement follows immediately. 

Let us now move to the third case. Here we have that:
\begin{align*}
|A_S|+ |D_S| \geq k \implies \min(|A_S|, k) + |D_S| \geq k,
\end{align*}
from which our statement follows immediately. 

Regarding the last case, say, first, that $|A_S|+k - m > \frac{|S|}{n}(|A_S|+|D_S|)$ (that is, the negated last condition in the fourth case). Using the previous observation that every $k$-element set $W$ provides to every voter $i\in S$ at least the satisfaction $|A_S|+k-m$, we immediately obtain that $u_i(W) \geq \frac{|S|}{n}(|A_S|+|D_S|)$.
Finally, we simply need to show that in the remaining part of the last case we have that $|A_S| \leq k$, or equivalently, that $|A_S| > k$ is covered by the previous cases. Indeed, if $|A_S| \geq k$ then all previous conditions that involve $A_S$ are satisfied. At the same time the conditions involving only $D_S$ cover the whole space of possibilities. 
\end{proof}

\section{Supplementary Material for the Asymmetric Setting}
\subsection{Omitted Proofs for Tax-Based Rules}
\tmes*
\begin{proof}
    For proving that 
    Tax-Phragm\'en satisfies PJPR consider an $\ell$-positively-cohesive group $S,$ as witnessed by $T\subseteq A_S,$ $|T|\geq \ell$.
Fix a candidate $c\in T$. Any candidate $c \in T$ costs $\frac{|A_c|}{|A_c|-|D_c|}.$ By the fact that $A_c\supseteq S$ we have the following:
\begin{align*}
    |A_c| |D_c|\geq |S|  |D_c|
    \iff 
    |A_c| (|S|-|D_c|)\leq |S|  (|A_c|-|D_c|).
\end{align*}
Therefore, the cost of $c$ is upper bounded by
$\frac{|S|}{|S|-|D_c|}\leq |S| \frac{k}{\ell n}$.
Hence, $\ell$ candidates from $T$ cost in total at most $|S| \frac{k}{n}$.
Thus,
voters in $S$ have participated in buying at least $\ell$ candidates in total, i.e., $|W\cap 
   \cup_{i\in S} A_i|\geq \ell$.

   Regarding the Group Veto axiom, we will show the following: Let $\mathcal{R}$ be a priceable rule such that $p(c)\geq \frac{|A_c|}{|A_c|-|D_c|}$ whenever $|A_c|>|D_c|,$ and $p(c) = \infty$, otherwise. Additionally, assume that the voters in $\mathcal{R}$ are initially endowed with the budget of at most $\nicefrac{k}{n}$. Then, $\mathcal{R}$ satisfies the Group Veto. Notice that both Tax-MES and Tax-Phragm\'en belong to the aforementioned family of rules.\looseness-1

   Consider a  candidate $c\in T$. 
  Since $A_c\subseteq ap(T),$ it holds that $|A_c|\leq |ap(T)|$ and since $S\subseteq D_c$ it holds that $|D_c|\geq |S|$, and so $|ap(T)|-|S|\geq |A_c|-|D_c|$. We get:
    \begin{gather*}
   \frac{1}{|S|}(|ap(T)|-|S|)\geq   
   \frac{1}{|D_c|}(|A_c|-|D_c|) \iff
    \frac{|ap(T)|}{|S|}-1\geq \frac{|A_c|}{|D_c|}-1
   \iff\\ 
     \frac{|D_c|}{|A_c|}\geq  \frac{|S|}{|ap(T)|} \iff
      1-\frac{|S|}{|ap(T)|}\geq 1-\frac{|D_c|}{|A_c|}\iff  \frac{|A_c|}{|A_c|-|D_c|}\geq \frac{|ap(T)|}{|ap(T)|-|S|}.
   \end{gather*}
We note that all fractions that appear above are well defined. This is simply because, by the definition of $\mathcal{R},$ it holds that $|D_c| < |A_c|$, and similarly $\frac{|S|}{|ap(T)|}\leq \frac{|D_c|}{|A_c|}<1$.

Since $p(c)\geq \frac{|A_c|}{|A_c|-|D_c|}$, we have that $p(c)\geq \frac{|ap(T)|}{|ap(T)|-|S|}$.
It holds that only voters from $ap(T)$ can pay for candidates from $T$ and jointly they have at most $|ap(T)|\frac{k}{n}$ money. Since the purchase of a candidate from $T$ costs at least $\frac{|ap(T)|}{|ap(T)|-|S|}$, they can afford a number of candidates that is at most 
$$\frac{|ap(T)|\frac{k}{n}}{\frac{|ap(T)|}{|ap(T)|-|S|}}=(|ap(T)|-|S|)\frac{k}{n}\leq \ell\cdot \frac{n}{k}\cdot \frac{k}{n}=\ell.$$ This implies in particular that at most $\ell$ candidates from $T$ are selected.
\end{proof}

\subsection{A Case Against Exhaustiveness and the Negative-Vote Incentives}\label{app:exhaustive}
In \Cref{sec:counterfactual}, we identified a limitation of the symmetric utility model arising from the presence of negative votes and we proposed steps to address it. Here, we discuss that the asymmetric setting also raises its own challenges.

First, one could wonder why the definition of tax-Phragm\'en we propose halts the procedure at time $\nicefrac{k}{n}$---a condition non-existent in the standard definition of Phragm\'en's rule. Without this stopping condition, the rule would always elect $k$ candidates (provided there are at least $k$ candidates with more in-favor than against votes). This is not always the case under tax-Phragm\'en, and this is intentional. An adaptation that wouldn't stop the procedure would fail to satisfy Group Veto. Nevertheless, it is still possible to formalize negative guarantees for the exhaustive variant.

Recall that the proof of \Cref{thm:mes} relies on voters having a limited budget. This, combined with the high cost of electing strongly opposed candidates, ensures that such candidates need significant support to be selected.
When voter funds are unlimited, the proposed rules could continue adding candidates as long as they have more supporters than opposers (respecting the committee size constraint). However, even in this case, vetoing preferences are still being respected. Candidates with less opposition are prioritized, and others are included only if no better options exist.
Below we formulate an axiom that sets the requirement on the level of support vetoed candidates must have in order to be elected.

\begin{definition}
    Consider an election $E=(C,V,k,B)$ in which there is a set of voters $S'$ such that $A_{S'}=\{c^*\},$ $D_{c^*}=\emptyset$. 
In such an election, given a positive integer $\ell \leq k$ and a set of at least $\ell$ candidates $T$, we say that a set of voters $S\subseteq V$ is weakly $(\ell,T)$-negatively-cohesive if $T\subseteq D_S$ and $|S|\geq |ap(T)|- \ell\cdot|S'|.$ An outcome $W$ provides weak Group Veto for $E$ if when $c^*\notin W$ then every weakly $(\ell,T)$-negatively-cohesive group satisfies $|W \cap T| < \ell$. A rule $\mathcal{R}$ satisfies the weak Group Veto axiom if for every election $E$ it outputs an outcome that provides weak Group Veto.
\hfill $\lrcorner$
\end{definition}
\begin{theorem}
Let $\mathcal{R}$ be a priceable rule such that $p(c)\geq \frac{|A_c|}{|A_c|-|D_c|}$ whenever $|A_c|>|D_c|,$ and $p(c) = \infty$, otherwise. Then, $\mathcal{R}$ satisfies weak Group Veto.
\end{theorem}
\begin{proof}
Let $S$ be a group of voters satisfying the conditions of weak Group Veto, specifically, there exists a set of candidates $T\subseteq \cap_{i\in S}D_i$ of cardinality $\ell$ and $|S| \geq |ap(T)|-\ell |S'|$.
Following the arguments from
the proof of \Cref{thm:mes}
we get that $p(c)\geq \frac{|ap(T)|}{|ap(T)|-|S|}$.

We now turn on computing $|W\cap T|$. Note that only voters from $ap(T)$ paid for including a candidate from $T$ into $W$. Voters from $ap(T)$ have spent at most $ap(T)\nicefrac{1}{|S'|}$ until the end of the execution of $\mathcal{R}$ as we will now justify. Say that until the end of the execution of the rule, each voter has been allocated and has been allowed to spent at most $q$. Voters from $S'$ didn't buy $c^*$ by assumption. Therefore, $|S'|\cdot q < 1 \Rightarrow q< \frac{1}{|S'|}.$ Hence, voters from $ap(T)$ had at most $ap(T)\nicefrac{1}{|S'|}$ to spend.
By the fact that each purchase cost at least $\frac{|ap(T)|}{|ap(T)|-|S|}$, voters in $ap(T)$ can afford a number of candidates that is strictly less than 
\begin{gather*}
\frac{|ap(T)|\frac{1}{|S'|}}{\frac{|ap(T)|}{|ap(T)|-|S|}}=(|ap(T)|-|S|)\frac{1}{|S'|}\leq \ell\cdot |S'|\frac{1}{|S'|}=\ell.
\end{gather*}
    This completes the proof.
\end{proof}
We note that the term ``exhaustive'' has been used in the committee elections literature for rules that always elect $k$ candidates. In our setting, fewer than $k$ candidates may be elected even with unlimited voter budgets. This occurs when fewer than $k$ candidates have more supporters than opposers.

On the other hand, tax-MES might select fewer than $k$ candidates even in the absence of negative votes---just as classic MES behaves~\citep{faliszewski2023participatory}. For that reason, classic MES is usually paired with a completion strategy to increase the number of elected candidates. However, in the presence of negative votes, pursuing exhaustiveness is less justified: adding a candidate can reduce satisfaction for some voters, so maximizing committee size is not always desirable. An approach to partially mitigate the issue of potentially having few elected candidates is to use a heuristic completion procedure after running tax-MES, adding candidates only if their inclusion does not violate the Group Veto axiom at the moment they are considered.

Another key characteristic in the asymmetric setting---particularly under the tax-based rules we propose---is that voting against a candidate incurs no cost, meaning that preventing a candidate $c$ from being selected does not diminish the voting power of those who vetoed $c$. By vetoing a neutral candidate, a voter effectively increases its cost, making it harder for that candidate's supporters to get it elected. This can result in the candidate being left out of the committee, thereby creating space for the voter's approved candidates.
Consequently, there is now a strong incentive for a voter to vote strategically against (all of their) neutral candidates. 
Nevertheless, we argue that such rules remain practical, especially when the number of negative votes per voter is limited. This restriction is common in real-world settings: for example, in most participatory budgeting elections, a voter is allowed to approve only a small number of projects---similarly, the election organizer could limit the number of negative votes expressed by a single voter. Another example is online platforms (such as Stack Overflow) which cap the number of downvotes a user can cast (per day).}

\end{document}